\documentclass[aps, prd, showkeys, showpacs ]{revtex4}

\usepackage[small]{caption}
\usepackage{amsmath,amsfonts,amscd,amssymb,epsf, epsfig, mathrsfs}\usepackage{graphicx}

\usepackage{lscape,rotating}

\newcommand{\pa}{\partial}

\newcommand{\vep}{\varepsilon}

\begin{document}

 \title{Material dependence of Casimir interaction between a   sphere and a   plate: First analytic correction beyond proximity force approximation}

\author{L. P. Teo}
 \email{LeePeng.Teo@nottingham.edu.my}
\affiliation{Department of Applied Mathematics, Faculty of Engineering, University of Nottingham Malaysia Campus, Jalan Broga, 43500, Semenyih, Selangor Darul Ehsan, Malaysia.}
\begin{abstract}We derive analytically the asymptotic behavior of the Casimir interaction between a sphere and a plate when the distance between them, $d$,  is much smaller than the radius of the sphere, $R$. The leading order and next-to-leading order terms are derived from the exact formula for the Casimir interaction energy. They are found to depend nontrivially on the dielectric functions of the objects. As expected, the leading order term coincides with that derived using the proximity force approximation. The result on the next-to-leading order term complements that found by Bimonte, Emig and Kardar [Appl. Phys. Lett. \textbf{100}, 074110 (2012)] using derivative expansion. Numerical results are presented when the dielectric functions are given by the plasma model or the Drude model, with the plasma frequency (for plasma and Drude models) and relaxation frequency (for Drude model) given respectively by 9eV and 0.035eV, the conventional values used for gold metal. It is found that if plasma model is used instead of Drude model, the error in the sum of the first two leading terms is at most 2\%, while the error in $\theta_1$, the ratio of the next-to-leading order term divided by $d/R$ to the leading order term,  can go up to 4.5\%.
\end{abstract}
\pacs{12.20.Ds,  11.10.-z}
\keywords{  Casimir interaction, sphere-plane configuration, analytic correction to proximity force approximation, plasma model, Drude model.}

\maketitle
\section{Introduction}
Casimir effect is a quantum effect that cannot be ignored in the realm of nanotechnology. It can cause malfunctions of nano devices due to stiction \cite{3,4,5}. In the last decade, intensive research have been carried out to determine the exact analytic formula for the Casimir effect between two nonplanar objects and its effective numerical computations (see, for example, the references cited in \cite{13}). Prior to this, one can only rely on the proximity force approximation (PFA) to compute an approximation for the Casimir interaction, and there is no way to determine the magnitude of the error in such an approximation.

In the  case of the sphere-plate setup, the most popular configuration used in Casimir experiments, there is only one curvature parameter given by the radius of the sphere, $R$. Hence it is expected that as $d$, the distance from the sphere to the plate, is much smaller than $R$, the Casimir interaction energy has an asymptotic expansion of the form
\begin{equation}\label{eq3_20_1}
E_{\text{Cas}}=E_{\text{Cas}}^{\text{PFA}}\left(1+\frac{d}{R}\theta_{1,E}+\ldots\right),
\end{equation}where $E_{\text{Cas}}^{\text{PFA}}$ is the proximity force approximation to the Casimir interaction energy. It follows that for the Casimir force $F_{\text{Cas}}$ and force gradient $\pa F_{\text{Cas}}/\pa d$, one also has expansions of the form
\begin{equation}\label{eq3_20_2}\begin{split}
F_{\text{Cas}}=F_{\text{Cas}}^{\text{PFA}}\left(1+\frac{d}{R}\theta_{1,F}+\ldots\right),
\\
\frac{\pa F_{\text{Cas}}}{\pa d}=\frac{\pa F_{\text{Cas}}^{\text{PFA}}}{\pa d}\left(1+\frac{d}{R}\theta_{1}+\ldots\right).
\end{split}\end{equation}
A few years ago, experiments have been set up to measure $\theta_1$ using a micromachined torsional oscillator \cite{14}. This gives a more ernest reason for the theoretical computation of the next-to-leading order terms of the Casimir interaction.
One of the breakthroughs in Casimir research brought by the achievement  in  explicit functional representation of the Casimir interaction  is that it becomes possible to compute analytically the next-to-leading terms, as has been shown in \cite{15,19,22} for the cylinder-plate configuration, in \cite{16,17,18} for the sphere-plate  configuration, in \cite{20} for the cylinder-cylinder configuration, and in \cite{21} for the sphere-sphere configuration. However, except for \cite{19}, all the other works only deal with ideal or non-physical boundary conditions, i.e., Dirichlet, Neumann, perfectly conducting, infinitely permeable or Robin boundary conditions. So far no work has discussed the exact analytical computation of the next-to-leading order term in the Casimir interaction between a sphere and a plate when both of these objects are made of real materials, and this is the goal of the current work to deal with this problem.

It should be mentioned that there has been an attempt to compute the material dependent next-to-leading order term in the Casimir interaction between a sphere and a plate carried out by Bimonte, Emig and Kadar \cite{12}, which used  the method of derivative expansion postulated in \cite{8}, which in turn is inspired by the work \cite{23}. However, it is still desirable to check the validity of the postulate in \cite{8,12} by computing the next-to-leading order terms from the exact formula for the Casimir interaction. Therefore, the results of our current work complement those obtained in \cite{12}.

\section{The Casimir interaction energy}
In this article, we recall the formula for the Casimir interaction between a sphere and a plate. Assume that the sphere  has relative permittivity $\vep_{r,1}$, and the plate has relative permittivity $\vep_{r,2}$.  When the thicknesses of the sphere and the plate are larger than their respective skin-depths, we can model this configuration by a ball and a semi-infinite space.   Let $d$ be the distance from the sphere to the plate, and let $L=d+R$, where $R$ is the radius of the ball.

  As shown in \cite{1,2}, the electromagnetic Casimir interaction energy of this sphere-plate configuration is given by
\begin{equation}\label{eq2_19_1}
E_{\text{Cas}}=\frac{\hbar}{2\pi}\int_0^{\infty} d\xi\text{Tr}\ln\left(1-\mathbb{M}(i\xi)\right),
\end{equation}
where the trace Tr is 
\begin{equation*}
\text{Tr}\, =\,\sum_{m=0}^{\infty}\sum_{l=\max\{1,|m|\}}^{\infty}\;\text{tr},
\end{equation*} with tr being the trace over $2\times 2$ matrices. The matrix elements of $\mathbb{M}$ are given by
\begin{equation}\label{eq3_19_6}
\begin{split}
\mathbb{M}_{lm,l'm'}=&\delta_{m,m'} \frac{(-1)^{m}\pi}{2}\sqrt{\frac{(2l+1)(2l'+1)}{l(l+1)l'(l'+1)}\frac{(l-m)!(l'-m)!}{(l+m)!(l'+m)!}}   \mathbb{T}^{lm}
 \int_{0}^{\infty}d\theta \sinh\theta e^{-2\kappa  L\cosh\theta}\\&
\times
 \left(\begin{aligned} \sinh\theta P_l^{m\prime}(\cosh\theta)\hspace{0.5cm} &-\frac{m}{\sinh\theta}P_l^m(\cosh\theta)\\ -\frac{m}{\sinh\theta}P_l^m(\cosh\theta)
 \hspace{0.4cm} & \quad\sinh\theta P_l^{m\prime}(\cosh\theta) \end{aligned}\right)\widetilde{\mathbb{T}}^{\theta}
 \left(\begin{aligned} \sinh\theta P_{l'}^{m'\prime}(\cosh\theta)\hspace{0.5cm} & \frac{m'}{\sinh\theta}P_{l'}^{m'}(\cosh\theta)\\  \frac{m'}{\sinh\theta}P_{l'}^{m'}(\cosh\theta)
 \hspace{0.4cm} & \quad\sinh\theta P_{l'}^{m'\prime}(\cosh\theta) \end{aligned}\right),
\end{split}\end{equation}where
\begin{equation*}
\mathbb{T}^{lm}=\begin{pmatrix} T_{lm}^{\text{TE}}& 0\\0 & T_{lm}^{\text{TM}}\end{pmatrix}
\end{equation*}is a diagonal matrix with elements
\begin{equation*}
\begin{split}
T_{lm}^{\text{TE}}(i\xi)=&
\frac{  I_{l+\frac{1}{2}}(\kappa R)\left(\frac{1}{2}I_{l+\frac{1}{2}}(n_1\kappa R)+n_1\kappa RI_{l+\frac{1}{2}}'(n_1\kappa R)\right)-
  I_{l+\frac{1}{2}}(n_1\kappa  R)\left(\frac{1}{2}I_{l+\frac{1}{2}}(\kappa R)+\kappa RI_{l+\frac{1}{2}}'(\kappa R)\right)}{  K_{l+\frac{1}{2}}(\kappa R)
\left(\frac{1}{2}I_{l+\frac{1}{2}}(n_1\kappa R)+n_1\kappa RI_{l+\frac{1}{2}}'(n_1\kappa R)\right)-
  I_{l+\frac{1}{2}}(n_1\kappa  R)\left(\frac{1}{2}K_{l+\frac{1}{2}}(\kappa R)+\kappa RK_{l+\frac{1}{2}}'(\kappa R)\right)},\\
T_{lm}^{\text{TM}}(i\xi)=&
\frac{  I_{l+\frac{1}{2}}(\kappa R)\left(\frac{1}{2}I_{l+\frac{1}{2}}(n_1\kappa R)+n_1\kappa RI_{l+\frac{1}{2}}'(n_1\kappa R)\right)-
\varepsilon_{r,1} I_{l+\frac{1}{2}}(n_1\kappa  R)\left(\frac{1}{2}I_{l+\frac{1}{2}}(\kappa R)+\kappa RI_{l+\frac{1}{2}}'(\kappa R)\right)}{  K_{l+\frac{1}{2}}
(\kappa R)\left(\frac{1}{2}I_{l+\frac{1}{2}}(n_1\kappa R)+n_1\kappa RI_{l+\frac{1}{2}}'(n_1\kappa R)\right)-
\varepsilon_{r,1} I_{l+\frac{1}{2}}(n_1\kappa  R)\left(\frac{1}{2}K_{l+\frac{1}{2}}(\kappa R)+\kappa RK_{l+\frac{1}{2}}'(\kappa R)\right)};
\end{split}\end{equation*}
and \begin{equation*}
\widetilde{\mathbb{T}}^{\theta}=\begin{pmatrix} \widetilde{T}_{\theta}^{\text{TE}}& 0\\0 &
\widetilde{T}_{\theta}^{\text{TM}}\end{pmatrix}
\end{equation*}is a diagonal matrix with elements
\begin{equation*}
\begin{split}
\widetilde{T}_{\theta}^{\text{TE}} =&\frac{ \sqrt{n_2^2+\sinh^2\theta}- \cosh\theta}
{ \sqrt{n_2^2+\sinh^2\theta}+ \cosh\theta},\\
\widetilde{T}_{\theta}^{\text{TM}} =&\frac{ \sqrt{n_2^2+\sinh^2\theta}-\vep_{r,2}\cosh\theta}
{ \sqrt{n_2^2+\sinh^2\theta}+\vep_{r,2}\cosh\theta}.
\end{split}
\end{equation*}Here \begin{equation*}
\kappa=\frac{\xi}{c}, \quad n_i=\sqrt{\varepsilon_{r,i}},\quad i=1,2,
\end{equation*}  and $P_l^m(x)$ are the associated Legendre functions given by
$$P_l^m(x)=\frac{(-1)^m}{2^ll!}(1-x^2)^{m/2}\frac{d^{l+m}}{dx^{l+m}}(x^2-1)^l$$
when $m\geq 0$, and
\begin{align}\label{eq3_19_5}P_l^{-m}(x)=(-1)^m\frac{(l-m)!}{(l+m)!}P_l^m(x).\end{align}
Direct numerical computations of the Casimir interaction energy from the formula \eqref{eq2_19_1} have been performed in a few works, for example, in \cite{24,10}. In numerical computations, the infinite matrix $\mathbb{M}$ has to be truncated to a matrix of finite size. A drawback of this direct numerical computation is that when $d/R$ gets smaller, one has to use a truncated matrix of larger size for accuracy, and this is subjected to the capacity of the computer. Currently, numerical computations are limited to $d/R>0.05$. However, in experiments, we usually have $d/R\sim 0.01$. Hence, analytical computation of the Casimir interaction energy becomes desirable.
\section{ Small separation asymptotic expansion  }\label{ae}In this section, we want to derive analytically the small separation asymptotic expansion of the Casimir interaction energy, Casimir force and the force gradient up to the next-to-leading order term.

One of the technical issues in the analytical computation of the Casimir interaction energy \eqref{eq2_19_1} is the appearance of the associated Legendre functions $P_l^m(x)$. First notice that because of the relation \eqref{eq3_19_5} and \begin{equation}\label{eq3_21_3}
\left(\begin{aligned} l-2k\hspace{0.5cm} &-\frac{m}{\sinh\theta} \\ -\frac{m}{\sinh\theta}
 \hspace{0.4cm} & \quad l-2k \end{aligned}\right)=\begin{pmatrix} 1 & 0\\ 0& -1\end{pmatrix}\left(\begin{aligned} l-2k\hspace{0.5cm} &\frac{m}{\sinh\theta} \\ \frac{m}{\sinh\theta}
 \hspace{0.4cm} & \quad l-2k \end{aligned}\right)\begin{pmatrix} 1 & 0\\ 0& -1\end{pmatrix},
\end{equation} the matrix element $\mathbb{M}_{lm,lm}$ \eqref{eq3_19_6} is equal to that when $m$ is changed to $-m$. Hence, it is sufficient to consider nonnegative $m$. In this case, one can show that
\begin{align*}
P_l^m(\cosh\theta)=&(-1)^mi^m\frac{(l+m)!}{ \pi l!}\int_{0}^{\pi}d\varphi \left(\cosh\theta+\sinh\theta\cos\varphi\right)^l \cos m\varphi\\
=&(-1)^mi^m\frac{(l+m)!}{\pi  }\sum_{k=0}^l\frac{1}{k!(l-k)!}e^{(l-2k)\theta}\int_{-\frac{\pi}{2}}^{\frac{\pi}{2}}d\varphi  \cos^{2l-2k}\varphi \sin^{2k}\varphi e^{2im\varphi}.\end{align*}Differentiating with respect to $\theta$ gives\begin{align*}
\sinh\theta P_l^{m\prime}(\cosh\theta)= &(-1)^mi^m\frac{(l+m)!}{\pi  }\sum_{k=0}^l\frac{l-2k}{k!(l-k)!}e^{(l-2k)\theta}\int_{-\frac{\pi}{2}}^{\frac{\pi}{2}}d\varphi  \cos^{2l-2k}\varphi \sin^{2k}\varphi e^{2im\varphi}.\end{align*}Therefore,
\begin{align*}
&\left(\begin{aligned} \sinh\theta P_l^{m\prime}(\cosh\theta)\hspace{0.5cm} &\frac{m}{\sinh\theta}P_l^m(\cosh\theta)\\ \frac{m}{\sinh\theta}P_l^m(\cosh\theta)
 \hspace{0.4cm} & \quad\sinh\theta P_l^{m\prime}(\cosh\theta) \end{aligned}\right)
\\ = &(-1)^mi^m\frac{(l+m)!}{\pi  }\sum_{k=0}^l\frac{1}{k!(l-k)!}
\left(\begin{aligned} l-2k\hspace{0.5cm} &\frac{m}{\sinh\theta} \\ \frac{m}{\sinh\theta}
 \hspace{0.4cm} & \quad l-2k \end{aligned}\right)e^{(l-2k)\theta}\int_{-\frac{\pi}{2}}^{\frac{\pi}{2}}d\varphi  \cos^{2l-2k}\varphi \sin^{2k}\varphi e^{2im\varphi}.
\end{align*}

Making a change of variables
$$ \frac{R\xi}{c}=\omega,   $$ and expanding the logarithm in \eqref{eq2_19_1}, we have
\begin{equation*}
E_{\text{Cas}}=-\frac{\hbar c}{2\pi R}\sum_{s=0}^{\infty}\frac{1}{s+1}\int_0^{\infty} d\omega \sum_{m=0}^{\infty}\left(\prod_{i=0}^s\sum_{l_i=\max{1,|m|}}^{\infty}\right)
\text{tr}\left(\prod_{i=0}^s\mathbb{M}_{l_im,l_{i+1}m}\right),
\end{equation*}where
\begin{equation}\label{eq2_20_1}\begin{split}
&\mathbb{M}_{l_im,l_{i+1}m}\\=& \frac{  1}{2\pi}\sqrt{ (2l_i+1)(2l_{i+1}+1) (l_i-m)!(l_{i+1}-m)! (l_i+m)!(l_{i+1}+m)! }  \begin{pmatrix} T_{l_i}^{\text{TE}}& 0\\0&-T_{l_i}^{\text{TM}}\end{pmatrix} \\&
\times\sum_{k=0}^{l_i}\sum_{k'=0}^{l_{i+1}}\frac{1}{k!(l_i-k)!}\frac{1}{k'!(l_{i+1}-k')!}\int_{0}^{\infty}d\theta \sinh\theta e^{-2\omega(1+\vep)\cosh\theta+(l_i+l_{i+1}-2k-2k')\theta}\\&\times\left(\begin{aligned} \frac{l_i-2k}{\sqrt{l_i(l_i+1)}}\hspace{1cm} & \frac{m}{\sinh\theta\sqrt{l_i(l_i+1)}} \\  \frac{m}{\sinh\theta\sqrt{l_i(l_i+1)}}
 \hspace{0.4cm} & \quad \frac{l_i-2k}{\sqrt{l_i(l_i+1)}} \end{aligned}\right) \begin{pmatrix} \widetilde{T}_{\theta}^{\text{TE}}& 0\\0&-\widetilde{T}_{\theta}^{\text{TM}}\end{pmatrix} \left(\begin{aligned} \frac{l_{i+1}-2k'}{\sqrt{l_{i+1}(l_{i+1}+1)}}\hspace{1cm} & \frac{m}{\sinh\theta\sqrt{l_{i+1}(l_{i+1}+1)}} \\  \frac{m}{\sinh\theta\sqrt{l_{i+1}(l_{i+1}+1)}}
 \hspace{0.4cm} & \quad \frac{l_{i+1}-2k'}{\sqrt{l_{i+1}(l_{i+1}+1)}} \end{aligned}\right)\\
 &\times \int_{-\frac{\pi}{2}}^{\frac{\pi}{2}}d\varphi  \cos^{2l_i-2k}\varphi \sin^{2k}\varphi e^{2im\varphi}\int_{-\frac{\pi}{2}}^{\frac{\pi}{2}}d\varphi'  \cos^{2l_{i+1}-2k'}\varphi' \sin^{2k'}\varphi' e^{2im\varphi'},
\end{split}\end{equation}with
\begin{equation*}
\begin{split}
T_{l_i}^{*} =&
\frac{ I_{l_i+\frac{1}{2}}(\omega)\left(\frac{1}{2}I_{l_i+\frac{1}{2}}(n_1\omega)+n_1\omega I_{l_i+\frac{1}{2}}'(n_1\omega)\right)-\alpha_1^*
 I_{l_i+\frac{1}{2}}(n_1\omega)\left(\frac{1}{2}I_{l_i+\frac{1}{2}}(\omega)+\omega I_{l_i+\frac{1}{2}}'(\omega)\right)}{ K_{l_i+\frac{1}{2}}(\omega)
\left(\frac{1}{2}I_{l_i+\frac{1}{2}}(n_1\omega)+n_1\omega I_{l_i+\frac{1}{2}}'(n_1\omega)\right)-\alpha_1^*
 I_{l_i+\frac{1}{2}}(n_1\omega)\left(\frac{1}{2}K_{l_i+\frac{1}{2}}(\omega)+\omega K_{l_i+\frac{1}{2}}'(\omega)\right)}.
\end{split}\end{equation*}Here $*$ = TE or TM, and $\alpha_1^{\text{TE}}=1$, $\alpha_1^{\text{TM}}=\vep_{r,1}$. The minus signs on $T_{l_i}^{\text{TM}}$ and $\widetilde{T}_{\theta}^{\text{TM}}$ in \eqref{eq2_20_1} come from the two matrices 
\begin{align*}
\begin{pmatrix} 1 & 0\\ 0&-1\end{pmatrix}
\end{align*}in \eqref{eq3_21_3}.
Let
$$e=\frac{d}{R}.$$
In the following, we make a shift of parameters
\begin{align*}
 l_i\mapsto l+l_i,\hspace{1cm}
\theta\mapsto \theta_0+\theta,
\end{align*}
where
\begin{align*}
l:=l_0,\hspace{1cm}\sinh\theta_0=\frac{l}{ \omega }.
\end{align*}When $e\ll 1$, the main contributions to the Casimir interaction energy come from the terms with
$$l\sim \frac{1}{e},\quad l_i\sim\frac{1}{\sqrt{e}},\quad m\sim\frac{1}{\sqrt{e}},\quad \omega\sim\frac{1}{e},\quad \theta\sim e.$$
In the small $e$ expansion below, we will count the order of $l, l_i, m, \omega$ and $ \theta$ as $1/e, 1/\sqrt{e}, 1/\sqrt{e}, 1/e$ and $e$ respectively.
Making a change of variables
\begin{align*}
\omega=\frac{l\sqrt{1-\tau^2}}{\tau},
\end{align*}we have
\begin{equation}\label{eq3_12_4}
E_{\text{Cas}}\approx-\frac{\hbar c}{2\pi R}\sum_{s=0}^{\infty}\frac{1}{s+1}\int_0^{1} \frac{1}{\tau^2\sqrt{1-\tau^2}}\int_0^{\infty} dl\;l \int_{-\infty}^{\infty} dm\left(\prod_{i=1}^s\int_{-\infty}^{\infty}dl_i\right)
\text{tr}\left(\prod_{i=0}^s\mathbb{M}_{(l+l_i)m,(l+l_{i+1})m}\right),
\end{equation}where $l_0=0$ and $l_{s+1}=0$ by default. The integration over $\theta$ is from $-\theta_0$ to $\infty$ which can be approximated by an integration from $-\infty$ to $\infty$, since $\theta$ is of order $e$ and $\theta_0$ is of order $1$.

Now we perform the small $e$ expansion of \eqref{eq2_20_1}. Writing $\cos\varphi$ as $\exp\left(-\ln\sec\varphi\right)$ and using the fact that
$$\ln\sec\varphi=\frac{\varphi^2}{2}+\frac{\varphi^4}{12},$$ we have the following small $e$   expansion:
\begin{equation}\label{eq3_21_1}\begin{split}
\int_{-\frac{\pi}{2}}^{\frac{\pi}{2}}d\varphi  \cos^{2(l+l_i)-2k}\varphi \sin^{2k}\varphi e^{2im\varphi}\approx & \int_{-\frac{\pi}{2}}^{\frac{\pi}{2}}d\varphi    \varphi^{2k}\left(1-\frac{\varphi^2}{6}\right)^{2k} \exp\left(-( l+l_i- k)\varphi^2-\frac{l+l_i-k}{6}\varphi^4\right)e^{2im\varphi}\\
\approx &\frac{1}{l^{k+1/2}}\int_{-\infty}^{\infty}d\varphi    \varphi^{2k}\left(1- \frac{k\varphi^2}{3l}\right)  \exp\left(-\frac{ l+l_i- k}{l}\varphi^2-\frac{l+l_i-k}{6 l^2}\varphi^4+\frac{2im\varphi}{\sqrt{l}}\right)\\
\approx &\frac{1}{l^{k+1/2}}\int_{-\infty}^{\infty}d\varphi    \varphi^{2k}\left(1+\mathcal{A}_{i,2}\right) \exp\left(\mathcal{B}_{i,1}+\mathcal{B}_{i,2}\right) \exp\left(-\varphi^2+ +\frac{2im\varphi}{\sqrt{l}}\right).
\end{split}\end{equation}In the second line, we have performed a rescaling $\varphi\mapsto \varphi/\sqrt{l}$ so that the main contribution to the integration over $\varphi$ comes from $\varphi$ that are $\sim 1$. Here and in the following, for any $\mathcal{X}$, $\mathcal{X}_{i,1}$ and $\mathcal{X}_{i,2}$ are,
respectively, terms of order $\sqrt{e}$ and $e$. When these terms do not depend on $i$, $i$ would be omitted.  Changing $l_i$ to $l_{i+1}$ and $k$ to $k'$ in \eqref{eq3_21_1}, we obtain a similar expansion:
\begin{align*}
\int_{-\frac{\pi}{2}}^{\frac{\pi}{2}}d\varphi'  \cos^{2(l+l_{i+1})-2k'}\varphi' \sin^{2k'}\varphi' e^{2im\varphi'}\approx &  \frac{1}{l^{k'+1/2}}\int_{-\infty}^{\infty}d\varphi'    \varphi^{\prime 2k}\left(1+\mathcal{C}_{i,2}\right) \exp\left(\mathcal{D}_{i,1}+\mathcal{D}_{i,2}\right) \exp\left(-\varphi^{\prime2}+\frac{2im\varphi'}{\sqrt{l}}\right).
\end{align*}Next, we can use Stirling's formula
$$\ln n!=\left(n+\frac{1}{2}\right)\ln n-n+\frac{1}{2}\ln 2\pi+\frac{1}{12 n}+\ldots $$to obtain an expansion
\begin{align*}
\frac{1}{l^{k+k'}}\frac{\sqrt{  (l+l_i-m)!(l+l_{i+1}-m)!(l+ l_i+m)!(l+l_{i+1}+m)! }  }{(l+l_i-k)!(l+l_{i+1}-k')!}\approx \exp\left(\frac{m^2}{l}+\mathcal{H}_{i,1}+\mathcal{H}_{i,2}\right).
\end{align*}On the other hand, we  have
\begin{align*}
 \frac{1}{2l} \sqrt{ (2l+2l_i+1)(2l+2l_{i+1}+1)}\approx  \left(1+\mathcal{G}_{i,1}+\mathcal{G}_{i,2}\right).
\end{align*}
For the terms  involving $\theta$, expanding in small $e$ gives
\begin{align*}
 \sinh(\theta+\theta_0)\approx \sinh\theta_0\left(1+\theta\coth\theta_0+\frac{\theta^2}{2}\right) \approx\frac{\tau}{\sqrt{1-\tau^2}}\left( 1+\mathcal{E}_{i,1}+\mathcal{E}_{i,2}\right);
\end{align*}
\begin{align*}
& \exp\Bigl(-2\omega(1+\vep)\cosh(\theta+\theta_0)+(2l+l_i+l_{i+1}-2k-2k')(\theta+\theta_0)\Bigr)\\\approx & \exp\Bigl((2l+l_i+l_{i+1}-2k-2k')\theta_0\Bigr) \\&\times\exp\left(-2\omega(1+\vep)\sinh\theta_0\left(\coth\theta_0+\theta+\frac{\theta^2}{2}\coth\theta_0+\frac{\theta^3}{6}
+\frac{\theta^4}{24}\coth\theta_0\right)+(2l+l_i+l_{i+1}-2k-2k')\theta\right)\\
\approx&\left(\frac{1-\tau}{1+\tau}\right)^{k+k'-\frac{l_i+l_{i+1}}{2}-l}    \exp\left(-\frac{2l}{\tau}- \frac{l\theta^2}{\tau}  - \frac{2el}{\tau}  +( l_i+l_{i+1})\theta+\mathcal{F}_{i,1}+\mathcal{F}_{i,2}\right);
\end{align*}
  \begin{align*}
&\left(\begin{aligned} \frac{l+l_i-2k}{\sqrt{(l+l_i)(l+l_i+1)}}\hspace{1cm} & \frac{m}{\sinh(\theta+\theta_0)\sqrt{(l+l_i)(l+l_i+1)}} \\  \frac{m}{\sinh(\theta+\theta_0)\sqrt{(l+l_i)(l+l_i+1)}}
 \hspace{0.4cm} & \quad \frac{l+l_i-2k}{\sqrt{(l+l_i)(l+l_i+1)}} \end{aligned}\right)\approx \begin{pmatrix} 1+ \mathcal{L}_{i,2} & \mathcal{M}_1 \\
 \mathcal{M}_1 &1 +\mathcal{L}_{i,2}\end{pmatrix},\\
 &\left(\begin{aligned} \frac{l+l_{i+1}-2k'}{\sqrt{(l+l_{i+1})(l+l_{i+1}+1)}}\hspace{1cm} & \frac{m}{\sinh(\theta+\theta_0)\sqrt{(l+l_{i+1})(l+l_{i+1}+1)}} \\  \frac{m}{\sinh(\theta+\theta_0)\sqrt{(l+l_{i+1})(l+l_{i+1}+1)}}
 \hspace{0.4cm} & \quad \frac{l+l_{i+1}-2k'}{\sqrt{(l+l_{i+1})(l+l_{i+1}+1)}} \end{aligned}\right)\approx \begin{pmatrix} 1 +\mathcal{N}_{i,2} & \mathcal{M}_1 \\
 \mathcal{M}_1  &1+ \mathcal{N}_{i,2}\end{pmatrix}.
\end{align*}
Here $$\mathcal{M}_1=\frac{m\sqrt{1-\tau^2}}{l\tau}$$ is of order $\sqrt{e}$. We do not need the term that is of order $e$ for the off-diagonal terms of  these matrices as they won't contribute to the next-to-leading order term of the Casimir interaction energy. Finally, the small $e$ expansions of $\widetilde{T}_{\theta+\theta_0}^{*}$ is the same as the small $\theta$ expansions:
\begin{equation*}
\begin{split}
\widetilde{T}_{\theta+\theta_0}^{*}=&\frac{ \sqrt{n_2^2+\sinh^2(\theta+\theta_0)}- \alpha_2^*\cosh(\theta+\theta_0)}
{ \sqrt{n_2^2+\sinh^2(\theta+\theta_0)}+\alpha_2^* \cosh(\theta+\theta_0)}\\
=&(-1)^{\text{sgn}^*}\widetilde{T}^{* }_{0}\left(1+  \theta\mathcal{K}^{*}_{1}+ \theta^2\mathcal{K}^{*}_{2}\right), 
\end{split}
\end{equation*}
where $\alpha_2^{\text{TE}}=1$, $\alpha_2^{\text{TM}}=\vep_{r,2}$, $\text{sgn}^{\text{TE}}=0, \text{sgn}^{\text{TM}}=1$,
 \begin{align*}
 \widetilde{T}^{\text{TE} }_{0}=&\frac{ \sqrt{ \vep_{r,2}(1-\tau^2)+\tau^2}-1}{\sqrt{\vep_{r,2}(1-\tau^2)+\tau^2}+1},\\
 \widetilde{T}^{\text{TM} }_{0}=& \frac{\vep_{r,2}- \sqrt{ \vep_{r,2}(1-\tau^2)+\tau^2} }{\vep_{r,2}+ \sqrt{ \vep_{r,2}(1-\tau^2)+\tau^2} },
 \\
\mathcal{K}^{\text{TE}}_{1}=& - \frac{2\tau}{\sqrt{\vep_{r,2}(1-\tau^2)+\tau^2}},\\
\mathcal{K}^{\text{TE}}_{2}=&  -\frac{\vep_{r,2}(1-\tau^2)}{(\vep_{r,2}(1-\tau^2)+\tau^2)^{3/2}}+\frac{2\tau^2}{\vep_{r,2}(1-\tau^2)+\tau^2},\\
\mathcal{K}^{\text{TM}}_{1}=&  \frac{2\vep_{r,2}\tau(1-\tau^2)}{\sqrt{\vep_{r,2}(1-\tau^2)+\tau^2}(\vep_{r,2}+\tau^2)},\\
\mathcal{K}^{\text{TM}}_{2}=& \frac{\vep_{r,2}^2(1-\tau^2)^2}{(\vep_{r,2}(1-\tau^2)+\tau^2)^{3/2}(\vep_{r,2}+\tau^2)}-\frac{\tau^2(-\vep_{r,2}^2\tau^2+\vep_{r,2}^2+\vep_{r,2}+1)}{(\vep_{r,2}(1-\tau^2)+\tau^2) (\vep_{r,2}+\tau^2)} \\
& +\frac{\tau^2\left(\vep_{r,2}\sqrt{\vep_{r,2}(1-\tau^2)+\tau^2}+1\right)^2}{\left(\vep_{r,2}(1-\tau^2)+\tau^2\right)\left(\sqrt{\vep_{r,2}(1-\tau^2)+\tau^2}+\vep_{r,2}\right)^2}.
\end{align*}Notice that $\widetilde{T}^{* }_{0},  \mathcal{K}_{ 1}^{*}, \mathcal{K}_{ 2}^{*}$ only depend on $\vep_{r,2}$ and $\tau$. They are independent of $l_i$ and $e$.

Gathering the expansions obtained above, we can write
\begin{align*}
&\mathbb{M}_{(l+l_i)m,(l+l_{i+1})m}\\\approx & \frac{1}{\pi}\begin{pmatrix} T_{l+l_i}^{\text{TE}}& 0\\0&-T_{l+l_i}^{\text{TM}}\end{pmatrix}\sum_{k=0}^{\infty}\frac{1}{k!}\sum_{k'=0}^{\infty}\frac{1}{k'!}\frac{\tau}{\sqrt{1-\tau^2}}\left(\frac{1-\tau}{1+\tau}\right)^{k+k'-\frac{l_i+l_{i+1}}{2}-l} \int_{-\infty}^{\infty}d\theta\exp\left(\frac{m^2}{l}-\frac{2l}{\tau}- \frac{l\theta^2}{\tau}  - \frac{2el}{\tau}  +( l_i+l_{i+1})\theta  \right)\\&\times \int_{-\infty}^{\infty}d\varphi    \varphi^{2k}   \exp\left(-\varphi^2+\frac{2im\varphi}{\sqrt{l}} \right) \int_{-\infty}^{\infty}d\varphi'    \varphi^{\prime 2k} \exp\left(-\varphi^{\prime2}+\frac{2im\varphi'}{\sqrt{l}} \right)   \left(1+\mathcal{O}_{i,1}+\mathcal{O}_{i,2} \right)\\&\times\begin{pmatrix} 1 +\mathcal{L}_{i,2} & \mathcal{M}_1 \\
 \mathcal{M}_1  &1+ \mathcal{L}_{i,2}\end{pmatrix} \begin{pmatrix}\widetilde{T}^{\text{TE} }_{0} \left(1+ \theta\mathcal{K}^{\text{TE}}_1  +\theta^2\mathcal{K}^{\text{TE}}_2\right)& 0\\ 0 &\widetilde{T}^{\text{TM} }_{0}\left(1+ \theta\mathcal{K}^{\text{TM}}_1  +\theta^2\mathcal{K}^{\text{TE}}_2\right)\end{pmatrix}  \begin{pmatrix} 1 +\mathcal{N}_{i,2} & \mathcal{M}_1 \\
 \mathcal{M}_1  &1+ \mathcal{N}_{i,2}\end{pmatrix},
\end{align*}where
\begin{align*}
&\exp\left(\mathcal{B}_{i,1}+\mathcal{B}_{i,2}+\mathcal{D}_{i,1}+\mathcal{D}_{i,2}+\mathcal{F}_{i,1}+\mathcal{F}_{i,2}+\mathcal{H}_{i,1}+\mathcal{H}_{i,2}
 \right) \left(1+\mathcal{A}_{i,2}\right) \left(1+\mathcal{C}_{i,2}\right)\left(1+\mathcal{E}_{i,1}+\mathcal{E}_{i,2}\right)\left(1+\mathcal{G}_{i,1}+\mathcal{G}_{i,2}\right) \\
\approx &1+\mathcal{O}_{i,1}+\mathcal{O}_{i,2}.
\end{align*}Notice that $\mathcal{M}_1, \mathcal{K}_1^{*}, \mathcal{K}_2^{*}$ are independent of $k, k', \varphi, \varphi'$ and $\theta$. Performing the summation over $k$ and $k'$ using the formulas
\begin{align*}
&\sum_{k=0}^{\infty} \frac{v^{k}}{k!}=e^{-v},\\
&\sum_{k=0}^{\infty} k\frac{v^{k}}{k!}=ve^{-v},\\
&\sum_{k=0}^{\infty} k^2\frac{v^{k}}{k!}=(v^2+v)e^{-v},
\end{align*}we obtain an expansion of the form
\begin{align*}
 \mathbb{M}_{(l+l_i)m,(l+l_{i+1})m} \approx &\frac{1}{\pi}\begin{pmatrix} T_{l+l_i}^{\text{TE}}& 0\\0&-T_{l+l_i}^{\text{TM}}\end{pmatrix} \frac{\tau}{\sqrt{1-\tau^2}}\left(\frac{1-\tau}{1+\tau}\right)^{ -\frac{l_i+l_{i+1}}{2}-l} \int_{-\infty}^{\infty}d\theta\exp\left(\frac{m^2}{l}-\frac{2l}{\tau}- \frac{l\theta^2}{\tau}  - \frac{2el}{\tau}  +( l_i+l_{i+1})\theta \right)\\&\times\int_{-\infty}^{\infty}d\varphi       \exp\left(-\frac{2\tau}{1+\tau}\varphi^2+\frac{2im\varphi}{\sqrt{l}} \right)  \int_{-\infty}^{\infty}d\varphi'    \exp\left(-\frac{2\tau}{1+\tau}\varphi^{\prime2}+\frac{2im\varphi'}{\sqrt{l}}  \right)   \left(1+\mathcal{P}_{i,1}+\mathcal{P}_{i,2}\right)\\
 &\times \begin{pmatrix}\widetilde{T}^{\text{TE} }_{0} \left(1+ \theta\mathcal{K}^{\text{TE}}_1+ \theta^2\mathcal{K}^{\text{TE}}_2+\mathcal{R}_2\right)+\widetilde{T}^{\text{TM} }_{0}\mathcal{M}_1^2& \left( \widetilde{T}_0^{\text{TE}}+\widetilde{T}_0^{\text{TM}}\right)\mathcal{M}_1\\ \left(\widetilde{T}_0^{\text{TE}}+\widetilde{T}_0^{\text{TM}}\right)\mathcal{M}_1 &\widetilde{T}^{\text{TM} }_{0}\left(1+ \theta\mathcal{K}^{\text{TM}}_1+ \theta^2\mathcal{K}^{\text{TM}}_2+\mathcal{R}_2\right)+\widetilde{T}^{\text{TE} }_{0}\mathcal{M}_1^2\end{pmatrix}.\end{align*}
 The $\mathcal{R}_2$ term comes from $\mathcal{L}_{i,2}$ and $\mathcal{N}_{i,2}$. The Gaussian integrations over $\varphi$ and $\varphi'$ can be performed straightforwardly and give
 \begin{align*}
 \mathbb{M}_{(l+l_i)m,(l+l_{i+1})m}   \approx &\frac{1}{2 }\begin{pmatrix} T_{l+l_i}^{\text{TE}}& 0\\0&-T_{l+l_i}^{\text{TM}}\end{pmatrix}  \frac{1+\tau}{\sqrt{1-\tau^2}}\left(\frac{1-\tau}{1+\tau}\right)^{ -\frac{l_i+l_{i+1}}{2}-l} \\&\times   \int_{-\infty}^{\infty}d\theta\exp\left(-\frac{m^2}{l\tau} -\frac{2l}{\tau}- \frac{l\theta^2}{\tau}  - \frac{2el}{\tau}  +( l_i+l_{i+1})\theta \right)\left(1+\mathcal{Q}_{i,1}+\mathcal{Q}_{i,2}\right)\\
 &\times \begin{pmatrix}\widetilde{T}^{\text{TE} }_{0} \left(1+\theta \mathcal{K}^{\text{TE}}_1+\theta^2 \mathcal{K}^{\text{TE}}_2+\mathcal{U}_2\right)+\widetilde{T}^{\text{TM} }_{0}\mathcal{M}_1^2& \left( \widetilde{T}_0^{\text{TE}}+\widetilde{T}_0^{\text{TM}}\right)\mathcal{M}_1\\ \left(\widetilde{T}_0^{\text{TE}}+\widetilde{T}_0^{\text{TM}}\right)\mathcal{M}_1 &\widetilde{T}^{\text{TM} }_{0}\left(1+ \theta\mathcal{K}^{\text{TM}}_1+ \theta^2\mathcal{K}^{\text{TM}}_2+\mathcal{U}_2\right)+\widetilde{T}^{\text{TE} }_{0}\mathcal{M}_1^2\end{pmatrix}.\end{align*}$\mathcal{U}_2$ comes from $\mathcal{R}_2$ and it is independent of $\theta$. Before performing integration over $\theta$,  one is supposed to multiply $\left(1+\mathcal{Q}_{i,1}+\mathcal{Q}_{i,2}\right)$ into the matrix after it. Up to the terms of order $e$, we  can write
 $$1+\mathcal{Q}_{i,1}+\mathcal{Q}_{i,2}\approx \left(1+\mathcal{Q}_{i,2}\right)\left(1+\mathcal{Q}_{i,1}\right),$$
 and only multiply $\left(1+\mathcal{Q}_{i,1}\right)$ into the matrix. On the other hand, up to the terms of order $e$, we can extract the term $\mathcal{U}_2$ of order $e$ out from the matrix. These give   \begin{align*}
&\mathbb{M}_{(l+l_i)m,(l+l_{i+1})m}   \\\approx &\frac{1}{2 }\begin{pmatrix} T_{l+l_i}^{\text{TE}}& 0\\0&-T_{l+l_i}^{\text{TM}}\end{pmatrix}  \left(\frac{1-\tau}{1+\tau}\right)^{ -\frac{l_i+l_{i+1}}{2}-l-\frac{1}{2}} \int_{-\infty}^{\infty}d\theta    \exp\left(-\frac{m^2}{l\tau} -\frac{2l}{\tau}- \frac{l\theta^2}{\tau}  - \frac{2el}{\tau}  +( l_i+l_{i+1})\theta \right)\left(1+\mathcal{Q}_{i,2}+\mathcal{U}_2\right)\\& \begin{pmatrix}\widetilde{T}^{\text{TE} }_{0} \left(1+\mathcal{Q}_{i,1}+\theta \mathcal{K}^{\text{TE}}_1 +\theta\mathcal{Q}_{i,1}\mathcal{K}^{\text{TE}}_1+\theta^2 \mathcal{K}^{\text{TE}}_2 \right)+\widetilde{T}^{\text{TM} }_{0}\mathcal{M}_1^2& \left( \widetilde{T}_0^{\text{TE}}+\widetilde{T}_0^{\text{TM}}\right)\mathcal{M}_1\\ \left(\widetilde{T}_0^{\text{TE}}+\widetilde{T}_0^{\text{TM}}\right)\mathcal{M}_1 &\widetilde{T}^{\text{TM} }_{0}\left(1+\mathcal{Q}_{i,1}+\theta \mathcal{K}^{\text{TM}}_1 +\theta\mathcal{Q}_{i,1}\mathcal{K}^{\text{TM}}_1+ \theta^2\mathcal{K}^{\text{TM}}_2 \right)+\widetilde{T}^{\text{TE} }_{0}\mathcal{M}_1^2\end{pmatrix}.\end{align*}Performing the Gaussian integration over $\theta$, we have
\begin{equation}\label{eq3_12_2}\begin{split}
&\mathbb{M}_{(l+l_i)m,(l+l_{i+1})m}   \\\approx&\frac{\sqrt{\pi\tau}}{2\sqrt{l} }\begin{pmatrix} T_{l+l_i}^{\text{TE}}& 0\\0&-T_{l+l_i}^{\text{TM}}\end{pmatrix}    \left(\frac{1-\tau}{1+\tau}\right)^{ -\frac{l_i+l_{i+1}}{2}-l-\frac{1}{2}}  \exp\left(-\frac{m^2}{l\tau} -\frac{2l}{\tau} - \frac{2el}{\tau}+\frac{\tau}{4l}(l_i+l_{i+1})^2\right) \left(1 +\mathcal{S}_{i,2}+\mathcal{U}_2\right)\\
 &\times \begin{pmatrix}\widetilde{T}^{\text{TE} }_{0} \left(1+\mathcal{S}_{i,1}+  \mathcal{V}^{\text{TE}}_{i,1} +\widetilde{\mathcal{S}}_{i,2}\mathcal{K}^{\text{TE}}_1+ \mathcal{V}^{\text{TE}}_{i,2} \right)+\widetilde{T}^{\text{TM} }_{0}\mathcal{M}_1^2& \left( \widetilde{T}_0^{\text{TE}}+\widetilde{T}_0^{\text{TM}}\right)\mathcal{M}_1\\ \left(\widetilde{T}_0^{\text{TE}}+\widetilde{T}_0^{\text{TM}}\right)\mathcal{M}_1 &\widetilde{T}^{\text{TM} }_{0}\left(1+\mathcal{S}_{i,1}+  \mathcal{V}^{\text{TM}}_{i,1} +\widetilde{\mathcal{S}}_{i,2}\mathcal{K}^{\text{TM}}_1+ \mathcal{V}^{\text{TM}}_{i,2}  \right)+\widetilde{T}^{\text{TE} }_{0}\mathcal{M}_1^2\end{pmatrix},
\end{split}\end{equation}where
\begin{align*}
\mathcal{S}_{i,j}=&\frac{\sqrt{l}}{\sqrt{\pi\tau}}\int_{-\infty}^{\infty}d\theta \exp\left(- \frac{l\theta^2}{\tau} +( l_i+l_{i+1})\theta \right)\mathcal{Q}_{i,j},\quad j=1,2,\\
\widetilde{\mathcal{S}}_{i,2}=&\frac{\sqrt{l}}{\sqrt{\pi\tau}}\int_{-\infty}^{\infty}d\theta \exp\left(- \frac{l\theta^2}{\tau} +( l_i+l_{i+1})\theta \right)\theta\mathcal{Q}_{i,1},
\\
\mathcal{V}^{*}_{i,1}=&  \frac{\tau}{2l}(l_i+l_{i+1})\mathcal{K}_1^{*},\\
\mathcal{V}^{*}_{i,2}=& \left(\frac{\tau}{2l}+\frac{\tau^2}{4l^2}(l_i+l_{i+1})^2\right)\mathcal{K}_2^{*}.
\end{align*}

Next we consider the small $e$ expansions of  $T_{l+l_i}^{*}$. Debye asymptotic expansions of modified Bessel functions say that:
\begin{align*}
&I_{\nu}(\nu z) \approx \frac{1}{\sqrt{2\pi \nu}}\frac{e^{\nu\eta(z)}}{(1+z^2)^{1/4}} \left(1+\frac{u_1(\tau (z))}{\nu}\right),\\
&\frac{1}{2}I_{\nu}(\nu z)+\nu zI_{\nu}'(\nu z)\approx  \frac{\sqrt{\nu}e^{\nu\eta(z)}(1+z^2)^{1/4}}{\sqrt{2\pi}}\left(1+ \frac{m_1(\tau (z))}{\nu }\right),\\
&K_{\nu}(\nu z) \approx \sqrt{\frac{\pi}{2 \nu}}\frac{e^{-\nu\eta(z)}}{(1+z^2)^{1/4}} \left(1-\frac{u_1(\tau (z))}{\nu}\right),\\
&\frac{1}{2}K_{\nu}(\nu z)+\nu zK_{\nu}'(\nu z)\approx- \sqrt{\frac{\pi\nu}{2}}  e^{-\nu\eta(z)}(1+z^2)^{1/4}\left(1- \frac{m_1(\tau (z))}{\nu }\right),
\end{align*}where
\begin{align*}
u_1(\tau )=&\frac{\tau }{8}-\frac{5\tau ^3}{24},\hspace{1cm}m_1(\tau )=\frac{\tau }{8}+\frac{7\tau ^3}{24},\\
\tau (z)=&\frac{1}{\sqrt{1+z^2}},\hspace{1cm} \eta(z)=\sqrt{1+z^2}+\ln\frac{z}{1+\sqrt{1+z^2}}.
\end{align*}
Let
\begin{align*}
z=\frac{\omega}{l+l_i+\frac{1}{2}},\quad z_1=n_1 z,\quad \nu=l+l_i+\frac{1}{2}.
\end{align*}
Then we find that up to terms of order $e$, we have
\begin{equation*}
\begin{split}
T_{l+l_i}^{*} \approx &\frac{e^{2\nu\eta(z)}}{\pi}
\frac{ \sqrt{1+z_1^2}\left(1+\frac{u_1(\tau (z))}{\nu}  +\frac{m_1(\tau (z_1))}{\nu }\right)-\alpha_1^*\sqrt{1+z^2}
    \left(1+\frac{u_1(\tau (z_1))}{\nu} + \frac{m_1(\tau (z))}{\nu }\right)}{\sqrt{1+z_1^2} \left(1-\frac{u_1(\tau (z))}{\nu} + \frac{m_1(\tau (z_1))}{\nu }\right)+\alpha_1^*
\sqrt{1+z^2} \left(1+\frac{u_1(\tau (z_1))}{\nu} - \frac{m_1(\tau (z))}{\nu }\right)}.
\end{split}\end{equation*}
In small $e$ expansion,
\begin{align*}
e^{2\nu\eta(z)} \approx &C^{l_i-l_{i+1}} \left(\frac{1-\tau}{1+\tau}\right)^{ \frac{l_i+l_{i+1}}{2}+l+\frac{1}{2}}    \exp\left( \frac{2l}{\tau}-\frac{\tau}{2l}(l_i^2+l_{i+1}^2)+\mathcal{I}_{i,1}+\mathcal{I}_{i,2}\right).\end{align*}
Therefore, we have an expansion of the form
\begin{equation}\label{eq3_12_1}\begin{split}
\begin{pmatrix} T_{l+l_i}^{\text{TE}}& 0\\0&-T_{l+l_i}^{\text{TM}}\end{pmatrix} \approx &\frac{C^{l_i-l_{i+1}}}{l} \left(\frac{1-\tau}{1+\tau}\right)^{ \frac{l_i+l_{i+1}}{2}+l+\frac{1}{2}}    \exp\left( \frac{2l}{\tau}-\frac{\tau}{2l}(l_i^2+l_{i+1}^2)+\mathcal{I}_{i,1}+\mathcal{I}_{i,2}\right)\\&\times\begin{pmatrix}T^{\text{TE} }_{0} \left(1+ \mathcal{J}^{\text{TE}}_{i,1}+ \mathcal{J}^{\text{TE}}_{i,2}\right)& 0\\ 0 &T^{\text{TM} }_{0}\left(1+ \mathcal{J}^{\text{TM}}_{i,1}+ \mathcal{J}^{\text{TM}}_{i,2}\right)\end{pmatrix},
\end{split}\end{equation} where
\begin{align*}
T^{\text{TE} }_{0}=&\frac{ \sqrt{ \vep_{r,1}(1-\tau^2)+\tau^2}-1}{\sqrt{\vep_{r,1}(1-\tau^2)+\tau^2}+1},\\
 T^{\text{TM} }_{0}=& \frac{\vep_{r,1}-\sqrt{\vep_{r,1}(1-\tau^2)+\tau^2} }{\vep_{r,1}+\sqrt{\vep_{r,1}(1-\tau^2)+\tau^2} },\\
\mathcal{J}_{i,1}^{*}=&  \frac{\tau l_i}{2l}\mathcal{W}_{ 1}^{*}\\
\mathcal{J}_{i,2}^{*} =&\frac{\tau^2l_i^2}{4l^2}\mathcal{W}_{ 2}^{*} +\frac{\tau}{l}\mathcal{Y}_{ 2}^{*},
\end{align*}with
\begin{align*}
\mathcal{W}_{ 1}^{\text{TE}}=&-\frac{4\tau}{\sqrt{\vep_{r,1}(1-\tau^2)+\tau^2}},\\
\mathcal{W}_{ 2}^{\text{TE}}=&\frac{ 8\tau^2+4\tau^4+4\vep_{r,1}-4\vep_{r,1}\tau^4}{ (\vep_{r,1}(1-\tau^2)+\tau^2)^{3/2}}+ \frac{4(1-\tau^2)^2\left(\vep_{r,1}+\sqrt{\vep_{r,1}(1-\tau^2)+\tau^2}\right)^2}{ \tau^2\left(\vep_{r,1}(1-\tau^2)+\tau^2\right)\left(\sqrt{\vep_{r,1}(1-\tau^2)+\tau^2}+1\right)^2}\\&-\frac{4(1-\tau^2)\left(\tau^2 + \vep_{r,1}\right)}{ \tau^2\left(\vep_{r,1}(1-\tau^2)+\tau^2\right)},\\
\mathcal{Y}_{ 2}^{\text{TE}}=&-\frac{\tau}{ (\vep_{r,1}(1-\tau^2)+\tau^2)^{1/2}}-\frac{8\vep_{r,1}\tau^2 - 3\vep_{r,1} - 5\vep_{r,1}\tau^4 + 9\tau^2 + 5\tau^4}{12 (\vep_{r,1} (1- \tau^2) + \tau^2)},\\
\mathcal{W}_{ 1}^{\text{TM}}=&\frac{4 \vep_{r,1} \tau(1-\tau^2) }{ \sqrt{\vep_{r,1}(1-\tau^2)+\tau^2}(\tau^2 + \vep_{r,1})},\\
\mathcal{W}_{ 2}^{\text{TM}}=&-\frac{\vep_{r,1}(1-\tau^2) (8\tau^2+4\tau^4+4\vep_{r,1}-4\vep_{r,1}\tau^4)}{ (\vep_{r,1}+\tau^2)(\vep_{r,1}(1-\tau^2)+\tau^2)^{3/2}}+ \frac{4(1-\tau^2)^2\vep_{r,1}^2\left(1+\sqrt{\vep_{r,1}(1-\tau^2)+\tau^2}\right)^2}{ \tau^2\left(\vep_{r,1}(1-\tau^2)+\tau^2\right)\left(\sqrt{\vep_{r,1}(1-\tau^2)+\tau^2}+\vep_{r,1}\right)^2} \\
& -\frac{4\vep_{r,1}^2(1-\tau^2)^3}{ \tau^2\left(\tau^2 + \vep_{r,1}\right)\left(\vep_{r,1}(1-\tau^2)+\tau^2\right)},\\
\mathcal{Y}_{ 2}^{\text{TM}}=&\frac{\vep_{r,1}(1-\tau^2)\tau}{ (\vep_{r,1}+\tau^2)(\vep_{r,1}(1-\tau^2)+\tau^2)^{1/2}}-\frac{7\vep_{r,1}^2\tau^4 - 4\vep_{r,1}^2\tau^2 - 3\vep_{r,1}^2 - 5\vep_{r,1}\tau^6 + 13\vep_{r,1}\tau^4 - 18\vep_{r,1}\tau^2 + 5\tau^6 - 3\tau^4}{12 (\vep_{r,1}+\tau^2)(\vep_{r,1} (1- \tau^2) + \tau^2)}.
\end{align*}Notice that $T^{* }_{0},  \mathcal{W}_{ 1}^{*}, \mathcal{W}_{ 2}^{*}, \mathcal{Y}_{ 2}^{*}$ only depend on $\vep_{r,1}$ and $\tau$. They are independent of $l_i$ and $e$.

Substituting \eqref{eq3_12_1} into \eqref{eq3_12_2}, we have an expansion of the form:
 \begin{equation}\label{eq3_12_3}\begin{split}
  \mathbb{M}_{(l+l_i)m,(l+l_{i+1})m}   \approx &\frac{\sqrt{\pi}}{2 }C^{l_i-l_{i+1}} \sqrt{\frac{\tau}{l}}\left(1+\mathcal{T}_{i,1}+\mathcal{T}_{i,2}+\mathcal{U}_2\right)\exp\left(-\frac{m^2}{l\tau} -\frac{ 2el}{\tau}  - \frac{\tau}{4l}( l_i-l_{i+1})^2 \right)\\&\times
   \begin{pmatrix}T^{\text{TE} }_{0}\widetilde{T}^{\text{TE} }_{0}\Lambda^{\text{TE}} +T^{\text{TE} }_{0}\widetilde{T}^{\text{TM} }_{0}\mathcal{M}_1^2 &T^{\text{TE} }_{0}\left( \widetilde{T}_0^{\text{TE}}+\widetilde{T}_0^{\text{TM}}\right)\mathcal{M}_1\\T^{\text{TM} }_{0}\left(\widetilde{T}_0^{\text{TE}}+\widetilde{T}_0^{\text{TM}}\right)\mathcal{M}_1&T^{\text{TM} }_{0}\widetilde{T}^{\text{TM} }_{0}\Lambda^{\text{TM}}+T^{\text{TM} }_{0}\widetilde{T}^{\text{TE} }_{0}\mathcal{M}_1^2\end{pmatrix},
 \end{split}\end{equation}where
 \begin{align*}
 \mathcal{T}_{i,1}=&\mathcal{I}_{i,1},\\
 \mathcal{T}_{i,2}=&\mathcal{I}_{i,2}+\mathcal{S}_{i,2}+\frac{1}{2}\mathcal{I}_{i,1}^2,\\
 \Lambda^{*}=&1+ \mathcal{J}^{*}_{i,1}+\mathcal{S}_{i,1}+ \mathcal{V}^{*}_{i,1}+ \mathcal{J}^{*}_{i,1}\mathcal{V}^{*}_{i,1}+ \mathcal{J}^{*}_{i,1}\mathcal{S}_{i,1}+ \mathcal{J}^{*}_{i,2} +\widetilde{\mathcal{S}}_{i,2}\mathcal{K}^{*}_1+ \mathcal{V}^{*}_{i,2}.
 \end{align*}
 Substituting \eqref{eq3_12_3} into \eqref{eq3_12_4}, and extracting terms up to order $e$, we have
 \begin{equation*}\begin{split}
E_{\text{Cas}}\approx &-\frac{\hbar c }{2\pi^{(s+3)/2} R}\sum_{s=0}^{\infty}\frac{1}{s+1}\frac{1}{2^{s+1}} \int_0^1 d\tau\frac{\tau^{(s+1)/2}}{\tau^2\sqrt{1-\tau^2}}\int_0^{\infty}dl\,l^{-(s-1)/2} \int_{-\infty}^{\infty} dm\left(\prod_{i=1}^s\int_{-\infty}^{\infty}dl_i\right)\\&\times
\exp\left(-\frac{m^2(s+1)}{l\tau} -\frac{ 2el(s+1)}{\tau}  - \frac{\tau}{4l}\sum_{i=0}^s( l_i-l_{i+1})^2 \right)\\&\times \left\{\sum_{* =\text{TE}, \text{TM}}\left[T^{* }_{0}\widetilde{T}^{*}_{0}\right]^{s+1}  \left(1+\sum_{i=0}^{s}\sum_{j=0}^s\mathcal{Z}_{i,1}\mathcal{Z}_{j,1}+\sum_{i=0}^s\mathcal{Z}_{i,2}+(s+1)\mathcal{U}_2\right)+ X\mathcal{M}_1^2
\right.\\
 &+\sum_{* =\text{TE}, \text{TM}}\left[T^{*}_{0}\widetilde{T}^{*}_{0}\right]^{s+1} \left( \sum_{i=0}^s\sum_{j=0}^s\mathcal{Z}_{i,1}\mathcal{J}_{j,1}^{*}+\sum_{i=0}^s\sum_{j\neq i}\mathcal{Z}_{i,1}\mathcal{V}_{j,1}^{*}+
\sum_{i=0}^s\mathcal{T}_{i,1}\mathcal{V}_{i,1}^{*}+\sum_{i=0}^s\widetilde{\mathcal{S}}_{i,2}\mathcal{K}_1^{*} \right.\\&\left.\left.\hspace{3cm}\sum_{i=0}^s\sum_{j=i+1}^s \mathcal{J}^{*}_{i,1}\mathcal{J}^{*}_{j,1} + \sum_{i=0}^s\sum_{j=i+1}^s\mathcal{V}^{*}_{i,1}\mathcal{V}^{*}_{j,1} + \sum_{i=0}^s \mathcal{J}^{*}_{i,1}\sum_{j= 0}^s\mathcal{V}^{*}_{j,1}+\sum_{i=0}^s \mathcal{J}^{*}_{i,2}+ \sum_{j=0}^s\mathcal{V}^{*}_{i,2}\right)\right\},
\end{split}\end{equation*}where
\begin{align*}
\mathcal{Z}_{i,1}=&\mathcal{T}_{i,1}+\mathcal{S}_{i,1},\\
\mathcal{Z}_{i,2}=&\mathcal{T}_{i,2}+\mathcal{T}_{i,1}\mathcal{S}_{i,1},
\end{align*}and
\begin{align*}
X=& (s+1)\left\{\left(T^{\text{TE} }_{0}\widetilde{T}_0^{\text{TM}}+T^{\text{TM} }_{0}\widetilde{T}_0^{\text{TE}}\right)
\frac{\left[T^{\text{TE} }_{0}\widetilde{T}^{\text{TE} }_{0}\right]^{s+1 }
-\left[T^{\text{TM} }_{0}\widetilde{T}^{\text{TM} }_{0}\right]^{s+1} }{T_0^{\text{TE}}\widetilde{T}_0^{\text{TE}}-
T_0^{\text{TM}}\widetilde{T}_0^{\text{TM}}}+2T^{\text{TE} }_{0}\widetilde{T}_0^{\text{TE}}T^{\text{TM} }_{0}\widetilde{T}_0^{\text{TM}} \frac{\left[T^{\text{TE} }_{0}\widetilde{T}^{\text{TE} }_{0}\right]^{s }
-\left[T^{\text{TM} }_{0}\widetilde{T}^{\text{TM} }_{0}\right]^{s} }{T_0^{\text{TE}}\widetilde{T}_0^{\text{TE}}-
T_0^{\text{TM}}\widetilde{T}_0^{\text{TM}}}\right\}.
\end{align*}We have omitted those terms of order $\sqrt{e}$   since they are odd in one of the $l_i$ and thus would give zero after integration with respect to $l_i$. It follows that
\begin{equation*}
E_{\text{Cas}}\approx E_{\text{Cas}}^0+E_{\text{Cas}}^1.
\end{equation*}$E_{\text{Cas}}^0$ is the leading order term that comes from   those terms of order $e^0$, and $E_{\text{Cas}}^1$ is the next-to-leading order term that comes from those terms of order $e$. 
$\vep_{r,1}$ and $\vep_{r,2}$ are functions of 
$$\xi=\frac{c}{R}\frac{l\sqrt{1-\tau^2}}{\tau}.$$They are independent of $l_i$ and $m$.
Performing the Gaussian integration over $l_i$, $1\leq i\leq s$, and $m$, we find that
\begin{equation}\label{eq3_13_3}\begin{split}
E_{\text{Cas}}^0= &-\frac{\hbar c }{4\pi R}\sum_{s=0}^{\infty}\frac{1}{(s+1)^2} \int_0^1 \frac{d\tau}{\tau\sqrt{1-\tau^2}}\int_0^{\infty}dl\,l  \exp\left(-\frac{ 2el(s+1)}{\tau}\right)
\sum_{* =\text{TE}, \text{TM}}\left[T^{* }_{0}\widetilde{T}^{*}_{0}\right]^{s+1},
\end{split}\end{equation}\begin{equation}\label{eq3_18_1}\begin{split}
 E_{\text{Cas}}^1  =&-\frac{\hbar c }{4\pi R}\sum_{s=0}^{\infty}\frac{1}{(s+1)^2} \int_0^1 \frac{d\tau}{\tau\sqrt{1-\tau^2}}\int_0^{\infty}dl\,l  \exp\left(-\frac{ 2el(s+1)}{\tau}\right) \Biggl\{\sum_{* =\text{TE}, \text{TM}}\left[T^{* }_{0}\widetilde{T}^{*}_{0}\right]^{s+1}  \left(\mathscr{A}+\mathscr{C}^{*}+\mathscr{D}^{*}\right)
+X\mathscr{B} \Biggr\}.
\end{split}\end{equation}The explicit formulas for $\mathscr{A}$, $\mathscr{B}$, $\mathscr{C}^{*}$ and $\mathscr{D}^{*}$ are given by
\begin{align*}
\mathscr{A}=&\frac{e^2l\tau}{3}\left((s+1)^3 +2(s+1)\right)+\frac{e}{3}\left((\tau^2-2)(s+1)^2-3\tau (s+1)+2\tau^2-1\right),\\
&+\frac{\tau^4+\tau^2-12}{12l\tau}(s+1)+\frac{(1+\tau)(1-\tau^2)}{2l\tau}-\frac{\tau(1-\tau^2)}{3l }\frac{1}{s+1},\\
\mathscr{B}=&\frac{ 1-\tau^2}{2l\tau (s+1)},\\
\mathscr{C}^{*}=& C_{V }  \mathcal{K}^{*}_1+C_J \mathcal{W}^{*}_1,\\
\mathscr{D}^{*}=&D_{VV}  \mathcal{K}^{*2}_1+D_{VJ}  \mathcal{K}^{* }_1
 \mathcal{W}^{*}_1+D_{JJ} \mathcal{W}^{*2}_1+\left(\frac{s+1}{2}\frac{\tau}{l}+D_V\right) \mathcal{K}^{*}_2 +   D_J \mathcal{W}^{*}_2+ (s+1)\frac{\tau}{l}\mathcal{Y}^{*}_2, 
\end{align*}with
\begin{align*}
C_{V}=&-\frac{e\tau}{3}\left((s+1)^3+2(s+1)\right)+\frac{1-\tau^2}{6l}(s+1)^2+\frac{\tau}{2l}(s+1)+\frac{1-4\tau^2}{12l},\\
C_J=&-\frac{e\tau}{6}\left((s+1)^3-(s+1)\right)+\frac{1}{12l}\left((s+1)^2-1\right),\\
D_{VV}=&\frac{\tau}{12l}\left((s+1)^3-2(s+1)^2+2(s+1)-1\right),\\
D_{JJ}=&\frac{\tau}{48l}\left((s+1)^3-2(s+1)^2-(s+1)+2\right),\\
D_{VJ}=&\frac{\tau}{12l}\left((s+1)^3- (s+1) \right),\\
D_{V}=&\frac{\tau}{6l}\left(2(s+1)^2-3(s+1)+1\right),\\
D_{J}=&\frac{\tau}{12l}\left( (s+1)^2- 1\right).
\end{align*}

Using the fact that $\widetilde{T}^{* }_{0}, T^{* }_{0}, \mathcal{K}_{ 1}^{*}, \mathcal{K}_{ 2}^{*},   \mathcal{W}_{ 1}^{*}, \mathcal{W}_{ 2}^{*}, \mathcal{Y}_{ 2}^{*}$ are independent of $e$, it is straightforward to take derivative with respect to $d$. For the Casimir force $$F_{\text{Cas}}=-\frac{\pa E_{\text{Cas}}}{\pa d},$$ we find that
\begin{equation*}
F_{\text{Cas}}\approx F_{\text{Cas}}^0+F_{\text{Cas}}^1,
\end{equation*}where $F_{\text{Cas}}^0$ and $F_{\text{Cas}}^1$ are respectively the leading order and next-to-leading order terms with
\begin{align*}
F_{\text{Cas}}^0= &-\frac{\hbar c }{2\pi R^2}\sum_{s=0}^{\infty}\frac{1}{s+1} \int_0^1 \frac{d\tau}{\tau^2\sqrt{1-\tau^2}}\int_0^{\infty}dl\,l^2  \exp\left(-\frac{ 2el(s+1)}{\tau}\right)
\sum_{* =\text{TE}, \text{TM}}\left[T^{* }_{0}\widetilde{T}^{*}_{0}\right]^{s+1},
\end{align*}\begin{align*}
F_{\text{Cas}}^1 =&-\frac{\hbar c }{2\pi R^2}\sum_{s=0}^{\infty}\frac{1}{s+1} \int_0^1 \frac{d\tau}{\tau^2\sqrt{1-\tau^2}}\int_0^{\infty}dl\,l^2  \exp\left(-\frac{ 2el(s+1)}{\tau}\right) \Biggl\{\sum_{* =\text{TE}, \text{TM}}\left[T^{* }_{0}\widetilde{T}^{*}_{0}\right]^{s+1}  \left(\widetilde{\mathscr{A}}+\widetilde{\mathscr{C}}^{*}+\mathscr{D}^{*}\right)
+X\mathscr{B} \Biggr\}.
\end{align*}Here
\begin{align*}
\widetilde{\mathscr{A}}=&\frac{e^2l\tau}{3}\left((s+1)^3 +2(s+1)\right)-\frac{e}{3}\left(2(s+1)^2+3\tau (s+1)+1\right),\\
&+\frac{-\tau^4+5\tau^2-12}{12l\tau}(s+1)+\frac{1+\tau-\tau^2}{2l\tau}-\frac{\tau }{6l }\frac{1}{s+1},\\
\widetilde{\mathscr{C}}^{*}=&\widetilde{ C_{V }}  \mathcal{K}^{*}_1+\widetilde{C_J} \mathcal{W}^{*}_1,\\
\widetilde{C_{V}}=&-\frac{e\tau}{3}\left((s+1)^3+2(s+1)\right)+\frac{1}{6l}(s+1)^2+\frac{\tau}{2l}(s+1)+\frac{1}{12l},\\
\widetilde{C_J}=&-\frac{e\tau}{6}\left((s+1)^3-(s+1)\right)+\frac{(1+\tau^2)}{12l}\left((s+1)^2-1\right).
\end{align*}
For the force gradient $\pa F_{\text{Cas}}/\pa d$, the leading order and next-to-leading order terms are
\begin{align*}
 \frac{\pa F_{\text{Cas}}^0}{\pa d} = &\frac{\hbar c }{ \pi R^3}\sum_{s=0}^{\infty} \int_0^1 \frac{d\tau}{\tau^3\sqrt{1-\tau^2}}\int_0^{\infty}dl\,l^3  \exp\left(-\frac{ 2el(s+1)}{\tau}\right)
\sum_{* =\text{TE}, \text{TM}}\left[T^{* }_{0}\widetilde{T}^{*}_{0}\right]^{s+1},
\end{align*}\begin{align*}
  \frac{\pa F_{\text{Cas}}^1}{\pa d}   =&\frac{\hbar c }{ \pi R^3}\sum_{s=0}^{\infty}  \int_0^1 \frac{d\tau}{\tau^3\sqrt{1-\tau^2}}\int_0^{\infty}dl\,l^3  \exp\left(-\frac{ 2el(s+1)}{\tau}\right) \Biggl\{\sum_{* =\text{TE}, \text{TM}}\left[T^{* }_{0}\widetilde{T}^{*}_{0}\right]^{s+1}  \left(\widehat{\mathscr{A}}+\widehat{\mathscr{C}}^{*}+\mathscr{D}^{*}\right)
+X\mathscr{B} \Biggr\},
\end{align*}where
\begin{align*}
\widehat{\mathscr{A}}=&\frac{e^2l\tau}{3}\left((s+1)^3 +2(s+1)\right)-\frac{e}{3}\left((2+\tau^2)(s+1)^2+3\tau (s+1)+1+2\tau^2\right),\\
&+\frac{-\tau^4+9\tau^2-12}{12l\tau}(s+1)+\frac{1+\tau-\tau^2+\tau^3}{2l\tau},\\
\widehat{\mathscr{C}}^{*}=&\widehat{ C_{V }}  \mathcal{K}^{*}_1+\widehat{C_J} \mathcal{W}^{*}_1,\\
\widehat{C_{V}}=&-\frac{e\tau}{3}\left((s+1)^3+2(s+1)\right)+\frac{1+\tau^2}{6l}(s+1)^2+\frac{\tau}{2l}(s+1)+\frac{1+4\tau^2}{12l},\\
\widehat{C_J}=&-\frac{e\tau}{6}\left((s+1)^3-(s+1)\right)+\frac{(1+2\tau^2)}{12l}\left((s+1)^2-1\right).
\end{align*}

Let us compare the leading order term to the proximity force approximation. The Casimir energy density between a pair of parallel dielectric plates with relative permittivities $\vep_{r,1}$ and $\vep_{r,2}$ is given by the Lifshitz's formula \cite{11}:
\begin{equation}\label{eq3_13_1}\begin{split}
\mathcal{E}^{\parallel}_{\text{Cas}}(d)=&\frac{\hbar c }{4\pi^2}\int_0^{\infty} d\kappa \int_{\kappa}^{\infty} dq\,q \sum_{* =\text{TE}, \text{TM}}\ln\left(1-r_1^{*}r_2^{*}e^{-2qd}\right),
\end{split}\end{equation}
where
\begin{align*}
r_i^{\text{TE}}=&\frac{\sqrt{(\vep_{r,i}-1)\kappa^2+q^2}-q}{\sqrt{(\vep_{r,i}-1)\kappa^2+q^2}+q},\\
r_i^{\text{TM}}=&\frac{\vep_{r,i}q-\sqrt{(\vep_{r,i}-1)\kappa^2+q^2}}{\vep_{r,i}q+\sqrt{(\vep_{r,i}-1)\kappa^2+q^2}}.
\end{align*}
The proximity force approximation to the Casimir interaction energy between a sphere and a plate with relative permittivities $\vep_{r,1}$ and $\vep_{r,2}$ is given by
\begin{equation}\label{eq3_13_2}\begin{split}
E^{\text{PFA}}_{\text{Cas}}=&2\pi R\int_d^{\infty}du\mathcal{E}^{\parallel}_{\text{Cas}}(u).
\end{split}
\end{equation}
Expanding the logarithm in \eqref{eq3_13_1} and substitute into \eqref{eq3_13_2}, we find that
\begin{equation*}\begin{split}
E^{\text{PFA}}_{\text{Cas}}=&-\frac{\hbar c R}{2\pi}\sum_{s=0}^{\infty}\frac{1}{s+1}\int_d^{\infty}du \int_0^{\infty} d\kappa \int_{\kappa}^{\infty} dq\,q  e^{-2q(s+1)u}\sum_{* =\text{TE}, \text{TM}}\left[r_1^{*}r_2^{*} \right]^{s+1}\\
 =&-\frac{\hbar c R}{4\pi}\sum_{s=0}^{\infty}\frac{1}{(s+1)^2} \int_0^{\infty} d\kappa \int_{\kappa}^{\infty} dq  e^{-2q(s+1)d}\sum_{* =\text{TE}, \text{TM}}\left[r_1^{*}r_2^{*} \right]^{s+1}.
 \end{split}
\end{equation*}
Now making a change of variables
\begin{equation*}
q=\frac{l}{R\tau}, \quad \kappa =\frac{l\sqrt{1-\tau^2}}{R\tau},
\end{equation*}we finally obtain
\begin{equation*}\begin{split}
E^{\text{PFA}}_{\text{Cas}}=& -\frac{\hbar c  }{4\pi R}\sum_{s=0}^{\infty}\frac{1}{(s+1)^2} \int_0^{1} \frac{d\tau}{\tau\sqrt{1-\tau^2}} \int_{0}^{\infty} dl\,l   \exp\left(-\frac{ 2el(s+1)}{\tau}\right)\sum_{* =\text{TE}, \text{TM}}\left[r_1^{*}r_2^{*} \right]^{s+1},
\end{split}\end{equation*}where
\begin{align*}
r_i^{\text{TE}}=&\frac{\sqrt{\vep_{r,i}(1-\tau^2)+\tau^2}-1}{\sqrt{\vep_{r,i}(1-\tau^2)+\tau^2}+1},\\
r_i^{\text{TM}}=&\frac{\vep_{r,i} -\sqrt{\vep_{r,i}(1-\tau^2)+\tau^2}}{\vep_{r,i} +\sqrt{\vep_{r,i}(1-\tau^2)+\tau^2}}.
\end{align*}
Compare to \eqref{eq3_13_3}, we find that our result for the leading order term agrees completely with the proximity force approximation.
\section{Plasma model}

In this section, we consider the special case where the dielectric permittivities of the sphere and the plate are described by the plasma model:
\begin{align*}
\vep_{r,i}(i\xi)=1+\frac{\omega_{p,i}^2}{\xi^2},
\end{align*}
where $\omega_{p,i}$ is the plasma frequency of the material.

Let
$$\omega_{d,i}=\frac{\omega_{p,i}d}{c}.$$ In terms of the variables $$t=\frac{el}{\tau}$$ and $\tau$, we have
\begin{align}\label{eq3_19_4}
\vep_{r,i}= 1+ \frac{\omega_{d,i}^2}{t^2(1-\tau^2)}.\end{align}

\begin{table}[h]\caption{\label{T1} The coefficients $\beta_{i,j}$.}

\begin{tabular}{||c|c|c||}
\hline
\hline
\hspace{0.5cm} $\beta$\hspace{0.5cm} & \hspace{0.1cm}exact value \hspace{0.1cm}& \hspace{0.5cm}numerical value \hspace{0.5cm}\\
 \hline
 &&\\
$\beta_{0,0}$ & 1 &1 \\
&&\\
 $\beta_{1,0}$ &$\displaystyle-\frac{4}{3}$ & $-1.3333$\\
 &&\\
$\beta_{2,0}$ & $\displaystyle \frac{9}{5}$& $1.8$\\
&&\\
$\beta_{1,1}$ & $\displaystyle \frac{18}{5}$& $3.6$ \\
 &&\\
$\beta_{3,0}$ & $\displaystyle- \frac{16}{7}+\frac{32}{735}\pi^2 $& $ -1.8560$\\
 &&\\
$\beta_{2,1}$ & $\displaystyle-\frac{48}{7}$& $-6.8571$ \\
&&\\
$\beta_{4,0}$ & $\displaystyle \frac{25}{9} - \frac{326}{1323}\pi^2$& $0.3458$ \\
&&\\
$\beta_{3,1}$ & $\displaystyle \frac{100}{9} - \frac{326}{1323}\pi^2$& $8.6791$\\
&&\\
$\beta_{2,2}$ &$\displaystyle \frac{50}{3}$ & $ 16.6667$\\
 &&\\
$\beta_{5,0}$ & $\displaystyle - \frac{36}{11}+\frac{1220}{1617}\pi^2 - \frac{379}{32340} \pi^4$&$ 3.0322$ \\
  &&\\
$\beta_{4,1}$ & $\displaystyle- \frac{180}{11}+\frac{2440}{1617}\pi^2 $& $-1.4707$\\
 &&\\
$\beta_{3,2}$ &$\displaystyle- \frac{360}{11}+ \frac{1220}{1617}\pi^2 $ & $-25.2808$\\
 &&\\
\hline
\hline
\end{tabular}\end{table}

\begin{table} \caption{\label{T2} The coefficients $\lambda_{i,j}$.}

\begin{tabular}{||c|c|c||}
\hline
\hline
\hspace{0.5cm} $\lambda$\hspace{0.5cm} & \hspace{0.5cm}exact value \hspace{0.5cm}& \hspace{0.5cm}numerical value \hspace{0.5cm}\\
 \hline
 &&\\
$\lambda_{0,0}$ & $\displaystyle- \frac{20}{\pi^2}+\frac{1}{3} $&$-1.6931$ \\
&&\\
 $\lambda_{1,0}$ &$\displaystyle \frac{56}{3} \frac{1}{\pi^2}- \frac{32}{45}$ & $ 1.1802$\\
 &&\\
 $\lambda_{0,1}$ &$\displaystyle \frac{56}{3} \frac{1}{\pi^2}- \frac{14}{45}$ & $  1.5802$\\
 &&\\
$\lambda_{2,0}$ & $\displaystyle  - \frac{398}{21}\frac{1}{\pi^2}+\frac{401}{315} $& $ -0.6473$\\
&&\\
$\lambda_{1,1}$ & $\displaystyle - \frac{796}{21}\frac{1}{\pi^2}+\frac{454}{315} $& $ -2.3993 $ \\
&&\\
 $\lambda_{0,2}$ &$\displaystyle - \frac{398}{21}\frac{1}{\pi^2}+\frac{113}{315} $ & $-1.5615 $\\
  &&\\
$\lambda_{3,0}$ & $\displaystyle  \frac{410}{21} \frac{1}{\pi^2}- \frac{37}{18}+ \frac{286}{6615}\pi^2 $& $  0.3493 $\\
 &&\\
$\lambda_{2,1}$ & $\displaystyle\frac{410}{7} \frac{1}{\pi^2}- \frac{26}{7} $& $    2.2202$ \\
&&\\
 $\lambda_{1,2}$ &$\displaystyle \frac{410}{7} \frac{1}{\pi^2}- \frac{16}{7}$ & $  3.6488$\\
 &&\\
 $\lambda_{0,3}$ &$\displaystyle\frac{410}{21} \frac{1}{\pi^2}- \frac{79}{126}+ \frac{1}{6615}\pi^2  $ & $1.3527 $\\
 &&\\
$\lambda_{4,0}$ & $\displaystyle -\frac{69824}{3465}\frac{1}{\pi^2} +\frac{35141}{10395} - \frac{28022}{99225}\pi^2 $& $  -1.4484$ \\
&&\\
$\lambda_{3,1}$ & $\displaystyle - \frac{279296}{3465}\frac{1}{\pi^2}+\frac{84176}{10395} - \frac{2774}{14175}\pi^2  $& $  -2.0007$\\
&&\\
$\lambda_{2,2}$ &$\displaystyle  - \frac{139648}{1155} \frac{1}{\pi^2}+ \frac{742}{99}+\frac{32}{11025}\pi^2 $ & $ -4.7269$\\
&&\\
 $\lambda_{1,3}$ &$\displaystyle - \frac{279296}{3465}\frac{1}{\pi^2}+\frac{43856}{10395}-\frac{46558}{1091475}\pi^2$ & $  -4.3690$\\
 &&\\
 $\lambda_{0,4}$ &$\displaystyle - \frac{69824}{3465}\frac{1}{\pi^2}+\frac{14981}{10395}  - \frac{11962}{1091475}\pi^2$ & $ -0.7087$\\
  &&\\
$\lambda_{5,0}$ & $\displaystyle\frac{ 26732}{1287}\frac{1}{\pi^2} - \frac{150368}{27027}+ \frac{4937399}{5675670}\pi^2 - \frac{1142}{63063}\pi^4  $&$   3.3627 $ \\
  &&\\
$\lambda_{4,1}$ & $\displaystyle \frac{133660}{1287}\frac{1}{\pi^2}- \frac{35026}{2079} + \frac{ 773884}{567567}\pi^2 $& $ 7.1324$\\
 &&\\
$\lambda_{3,2}$ &$\displaystyle \frac{ 267320}{1287}\frac{1}{\pi^2}-\frac{548024}{27027} + \frac{26212}{51597}\pi^2   $ & $ 5.7822$\\
&&\\
 $\lambda_{2,3}$ &$\displaystyle \frac{267320}{1287}\frac{1}{\pi^2} -\frac{ 415724}{27027}+ \frac{16826}{81081}\pi^2  $ & $ 7.7116 $\\
&&\\
 $\lambda_{1,4}$ &$\displaystyle\frac{133660}{1287}\frac{1}{\pi^2} -\frac{256888}{27027}+ \frac{19984}{81081}\pi^2 $ & $  3.4503$\\
  &&\\
 $\lambda_{0,5}$ &$\displaystyle \frac{26732}{1287}\frac{1}{\pi^2} - \frac{84218}{27027}+ \frac{3329}{62370}\pi^2 + \frac{8059}{2522520}\pi^4 $ & $ -0.1736$\\
  &&\\
\hline
\hline
\end{tabular}
 \end{table}

First consider the case
$$\omega_{d,i} \gg 1.$$
 The limit  where $\omega_{d,i}\rightarrow\infty, i=1,2,$ is the perfect conductor limit. We can compute analytically the asymptotic expansion of the leading and next-to-leading order terms in the small parameters
$$a_i=\frac{1}{\omega_{d,i}}.$$ Specifically, we have
\begin{align}\label{eq3_19_1}
E_{\text{Cas}} \approx &-\frac{\pi^3\hbar c  R}{720 d^2}\left(\sum_{i=0}^{\infty}\sum_{j=0}^{\infty} \beta_{i,j}  a_1^ia_2^j+\frac{d}{R}\sum_{i=0}^{\infty}\sum_{j=0}^{\infty} \lambda_{i,j}  a_1^ia_2^j+\ldots\right),
\end{align}where $\beta_{0,0}=1$ and
\begin{align*}
E_{\text{Cas}}^{\text{PFA,PC}}= -\frac{\pi^3\hbar c  R}{720 d^2}
\end{align*}is the leading order approximation to the Casimir interaction energy between a perfectly conducting sphere and a perfectly conducting plate.
The exact values of $\beta_{i,j}$ and $\lambda_{i,j}$ for $i+j\leq 5$ are listed in  Table \ref{T1} and Table \ref{T2}. From \eqref{eq3_13_3}, it is obvious that the leading term is symmetric when we interchange $\vep_{r,1}$ with $\vep_{r,2}$. It follows that
$$\beta_{i,j}=\beta_{j,i}\quad\text{for all}\;(i,j).$$Hence, we only list the coefficients of $\beta_{i,j}$ when $i\geq j$ in Table \ref{T1}.

From \eqref{eq3_19_1}, we have
\begin{align}\label{eq3_19_2}
F_{\text{Cas}} \approx &-\frac{\pi^3\hbar c  R}{360 d^3}\left(\sum_{i=0}^{\infty}\sum_{j=0}^{\infty} \frac{(i+j+2)}{2}\beta_{i,j}  a_1^ia_2^j+\frac{d}{R}\sum_{i=0}^{\infty}\sum_{j=0}^{\infty} \frac{(i+j+1)}{2} \lambda_{i,j}  a_1^ia_2^j+\ldots\right),
\end{align}
\begin{align}\label{eq3_19_3}
\frac{\pa F_{\text{Cas}}}{\pa d} \approx &\frac{\pi^3\hbar c  R}{120 d^4}\left(\sum_{i=0}^{\infty}\sum_{j=0}^{\infty} \frac{(i+j+2)(i+j+3)}{6}\beta_{i,j}  a_1^ia_2^j+\frac{d}{R}\sum_{i=0}^{\infty}\sum_{j=0}^{\infty} \frac{(i+j+1)(i+j+2)}{6} \lambda_{i,j}  a_1^ia_2^j+\ldots\right).
\end{align}
 Here
\begin{align*}
F_{\text{Cas}}^{\text{PFA,PC}}= -\frac{\pi^3\hbar c  R}{360 d^3}
\end{align*}and
\begin{align*}
\frac{\pa F_{\text{Cas}}^{\text{PFA,PC}}}{\pa d}= \frac{\pi^3\hbar c  R}{120 d^4}
\end{align*}
are respectively the leading order approximations to the Casimir   force and force gradient between a perfectly conducting sphere and a perfectly conducting plate. Setting $a_1=a_2=0$ in \eqref{eq3_19_1}, \eqref{eq3_19_2} and \eqref{eq3_19_3}, we obtain
\begin{align*}
E_{\text{Cas}} \approx  &
E_{\text{Cas}}^{\text{PFA,PC}}\left(1+\frac{d}{R}\left[\frac{1}{3}-\frac{20}{\pi^2}\right]+\ldots\right),
\\
F_{\text{Cas}} \approx  &
F_{\text{Cas}}^{\text{PFA,PC}}\left(1+\frac{d}{R}\left[\frac{1}{6}-\frac{10}{\pi^2}\right]+\ldots\right),\\
\frac{\pa F_{\text{Cas}}}{\pa d} \approx   &
\frac{\pa F_{\text{Cas}}^{\text{PFA,PC}}}{\pa d}\left(1+\frac{d}{R}\left[\frac{1}{9}-\frac{20}{3\pi^2}\right]+\ldots\right),
\end{align*}which are well-known results for the leading and next-to-leading order terms of the perfectly conducting sphere-plate configuration \cite{18,8}.

\begin{figure}[h]
\epsfxsize=0.45\linewidth \epsffile{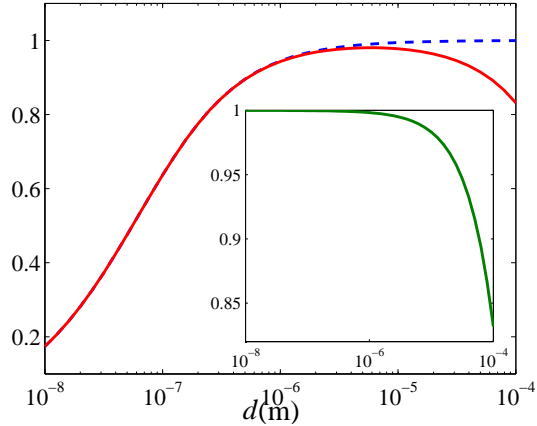} \caption{\label{f1} The   leading order term of the Casimir interaction energy  normalized by $E_{\text{Cas}}^{\text{PFA,PC}}$ (dashed line) and the sum of the leading and next-to-leading order terms normalized by $ E_{\text{Cas}}^{\text{PFA,PC}}$ (solid line). Inset is the ratio of the latter to the former.}\end{figure}

\begin{figure}[h]
\epsfxsize=0.45\linewidth \epsffile{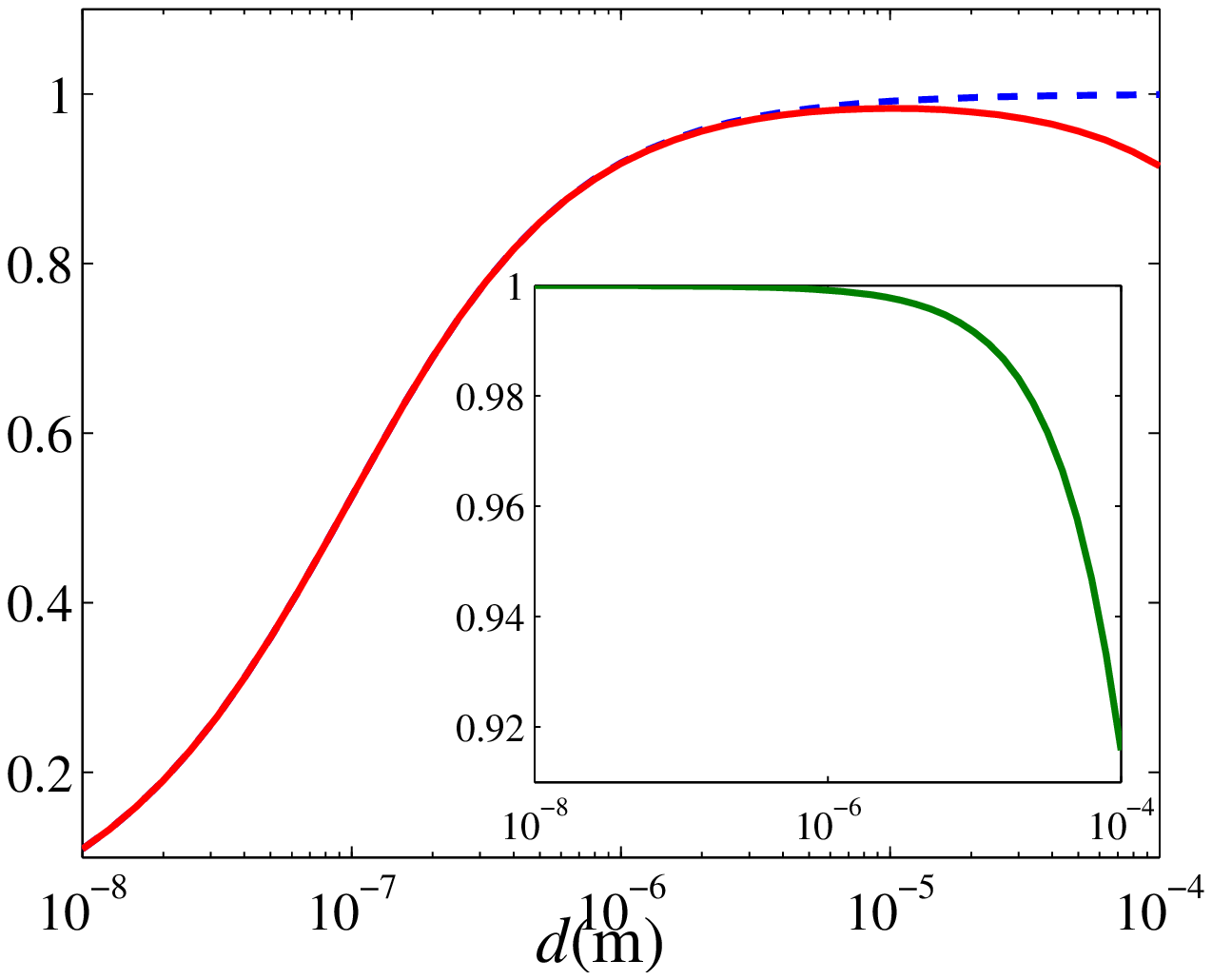} \caption{\label{f2} The   leading order term of the Casimir force  normalized by $F_{\text{Cas}}^{\text{PFA,PC}}$ (dashed line) and the sum of the leading and next-to-leading order terms normalized by $ F_{\text{Cas}}^{\text{PFA,PC}}$ (solid line). Inset is the ratio of the latter to the former.}\end{figure}

\begin{figure}[h]
\epsfxsize=0.45\linewidth \epsffile{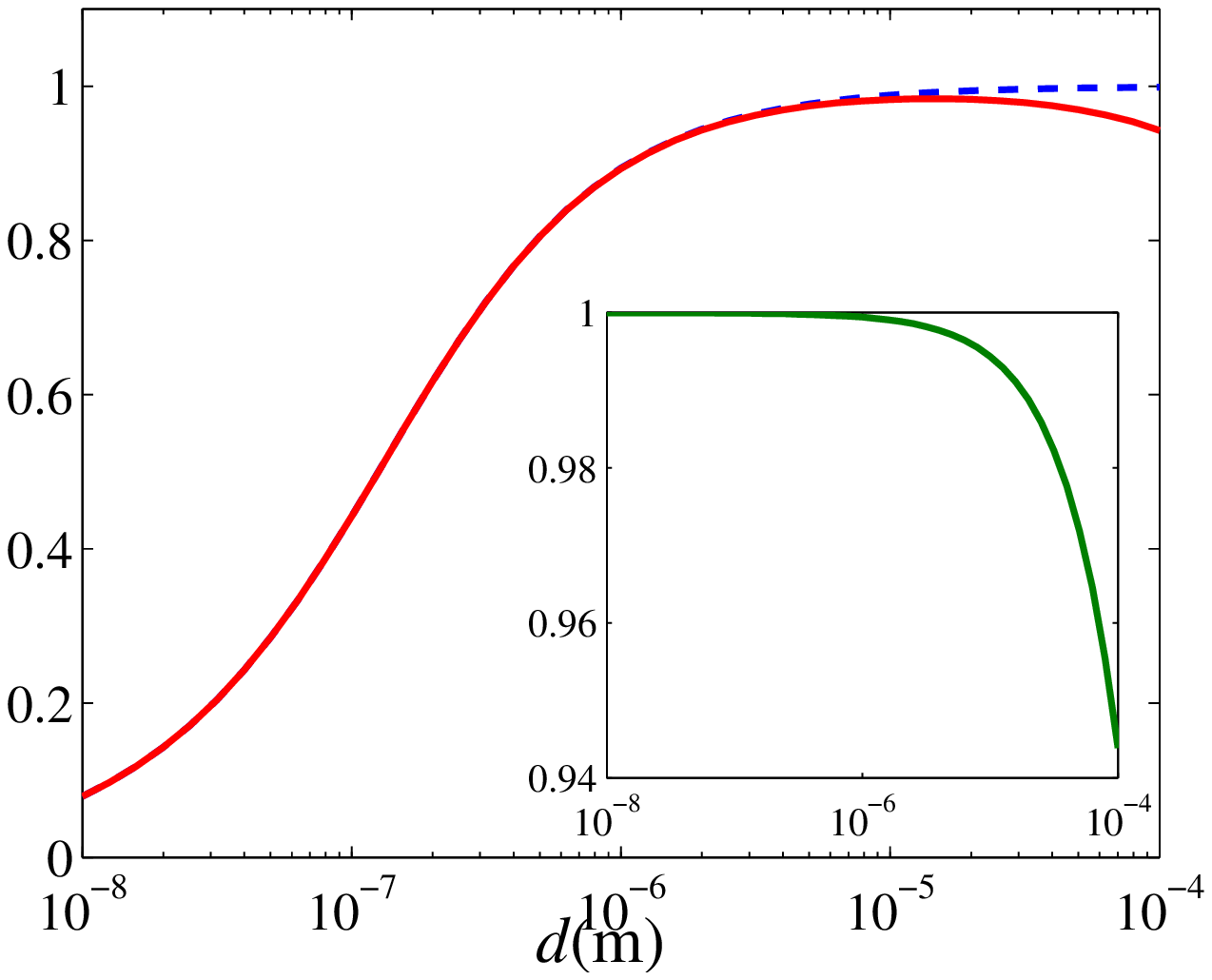} \caption{\label{f3} The   leading order term of the force gradient  normalized by $ \pa F_{\text{Cas}}^{\text{PFA,PC}}/\pa d$ (dashed line) and the sum of the leading and next-to-leading order terms normalized by $  \pa F_{\text{Cas}}^{\text{PFA,PC}}/\pa d$ (solid line). Inset is the ratio of the latter to the former.}\end{figure}

Next we consider numerical results with  $\omega_{p,1}=\omega_{p,2}=$ 9eV, which is the plasma frequency for gold \cite{6}.   The radius of the sphere $R$ is taken to be 1mm.
Substituting \eqref{eq3_19_4} into the formulas obtained in the previous section, we can compute numerically the leading order term (the proximity force approximation) and the next-to-leading order term of the Casimir interaction. In Figs. \ref{f1}, \ref{f2} and \ref{f3}, we plot the leading order term, the sum of the leading order and next-to-leading order terms of the Casimir interaction energy, Casimir force and force gradient, normalized respectively by $ E_{\text{Cas}}^{\text{PFA,PC}}, F_{\text{Cas}}^{\text{PFA,PC}}$ and $\pa F_{\text{Cas}}^{\text{PFA,PC}}/\pa d$, as a function of $d$ for $d$ between 10nm and 100$\mu$m. From the figures, we notice that when $d/R\sim 0.1$, the  corrections to PFA become significant and they would contribute corrections of about 10\%.



To have a better picture about the corrections to the proximity force approximations, define $\theta_{1,E}$, $\theta_{1,F}$ and $\theta_{1}$ by
\begin{align*}
\theta_{1,E}=&\frac{R}{d}\frac{E_{\text{Cas}}^1}{E_{\text{Cas}}^0},\\
\theta_{1,F}=&\frac{R}{d}\frac{F_{\text{Cas}}^1}{F_{\text{Cas}}^0},\\
\theta_{1}=&\frac{R}{d}\frac{ \pa F_{\text{Cas}}^1/\pa d}{ \pa F_{\text{Cas}}^0/\pa d },
\end{align*}so that \eqref{eq3_20_1} and \eqref{eq3_20_2} hold.
\begin{figure}
\epsfxsize=0.32\linewidth \epsffile{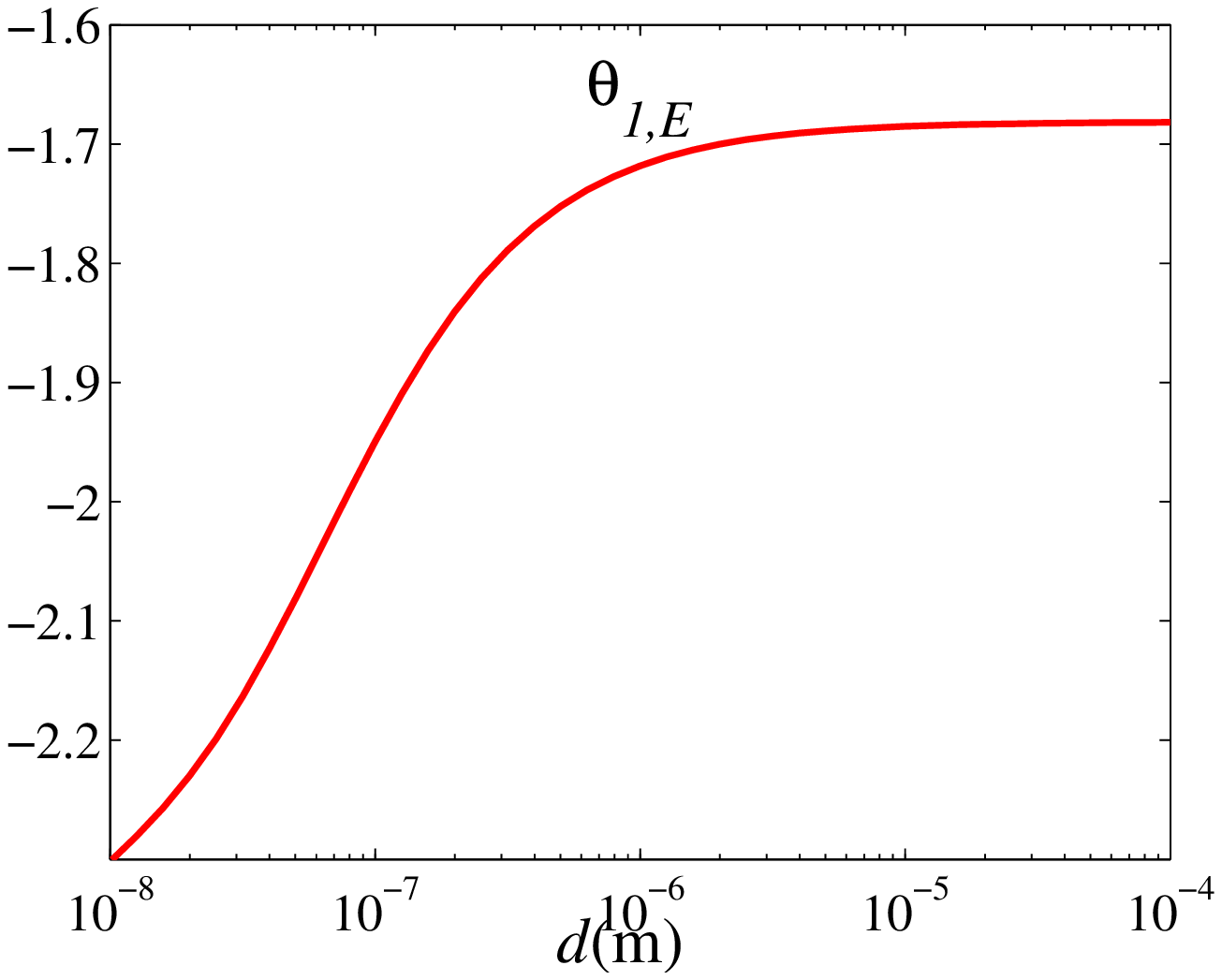} \epsfxsize=0.32\linewidth \epsffile{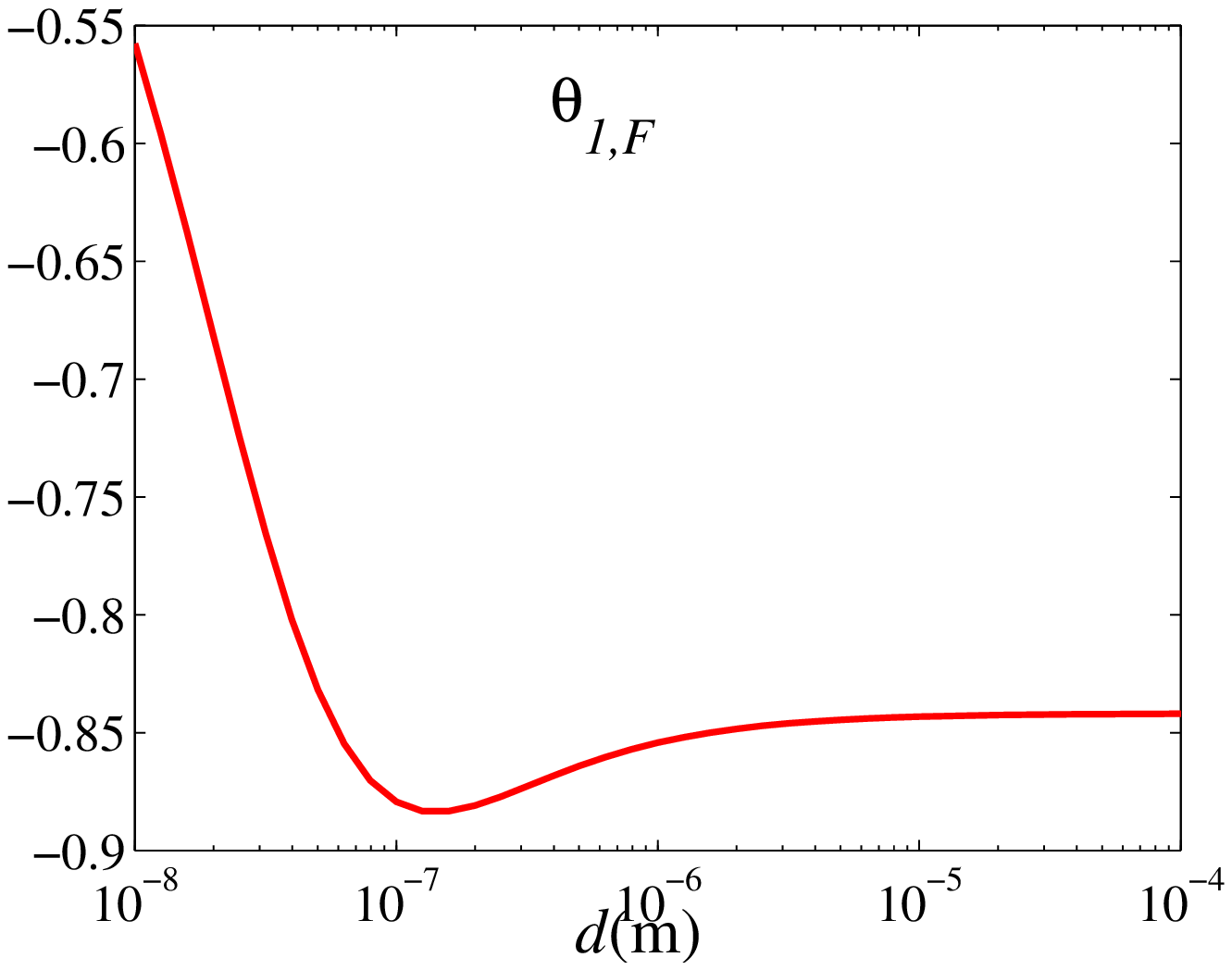} \epsfxsize=0.32\linewidth \epsffile{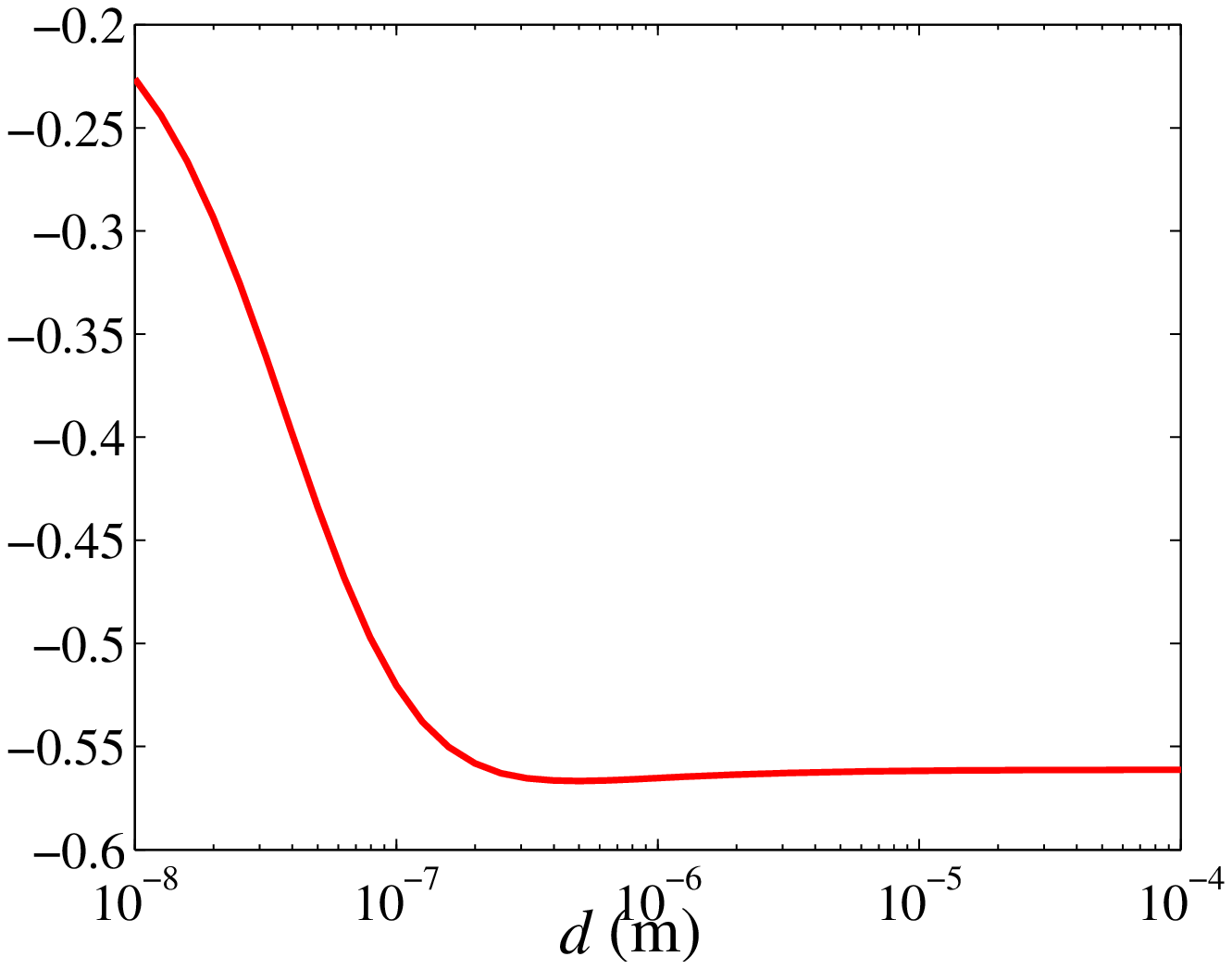} \caption{\label{f4}  $\theta_{1,E}(d)$,  $\theta_{1,F}(d)$ and $\theta_1(d)$ computed using plasma model. }\end{figure}

In Figs. \ref{f4}, we plot $\theta_{1,E}$, $\theta_{1,F}$ and $\theta_{1}$  as   functions of $d$ for $d$ between 10nm and 100$\mu$m.
As $d$ increases to 100$\mu$m, we find that $\theta_{1,E}$, $\theta_{1,F}$ and $\theta_{1}$ tend respectively to the values $\displaystyle 1/3-20/\pi^2 =-1.6931$, $\displaystyle  1/6-10/\pi^2=-0.8465 $ and $\displaystyle  1/9-20/(3\pi^2)=-0.5644 $, which are corresponding values for perfect conductors. This is not surprising since $\omega_{d,i}=\omega_{p,i}d/c$ is approximately equal to 5000 when $d=10^{-4}$. When $d$ is small, the deviations from the limiting values for perfect conductors are very significant. On the other hand, we also notice that   $\theta_{1,F}$ and $\theta_1$ are bounded below. $\theta_1$ is a quantity that can be measured experimentally \cite{14}. From Fig. \ref{f4}, we find that it is bounded below by $-0.57$.

\section{Drude model}
 
  The Drude dielectric function is given by
\begin{align*}
\vep_{r,i}(i\xi)=1+\frac{\omega_{p,i}^2}{\xi(\xi+\gamma_i)},
\end{align*} where $\gamma_i$ is the relaxation frequency of the material. In the limit where $\gamma_i\rightarrow 0$, the Drude dielectric function becomes the plasma dielectric function.

Let
$$\gamma_{d,i}=\frac{\gamma_{i}d}{c}.$$ In terms of the variables $t$ and $\tau$, we have
\begin{align}\label{eq3_19_4_2}
\vep_{r,i}= 1+ \frac{\omega_{d,i}^2}{t\sqrt{1-\tau^2}\left(t\sqrt{1-\tau^2}+\gamma_{d,i}\right)}.\end{align}
Substituting this into the results obtained in Section \ref{ae}, we can compute numerically the leading order and next-to-leading order terms of the Casimir interaction for Drude models.

Let us consider the case where $\omega_{p,1}=\omega_{p,2}=$ 9eV and $\gamma_1=\gamma_2= 0.035$ eV, which are the conventional values used for gold \cite{6}.

\begin{figure}[h]
\epsfxsize=0.45\linewidth \epsffile{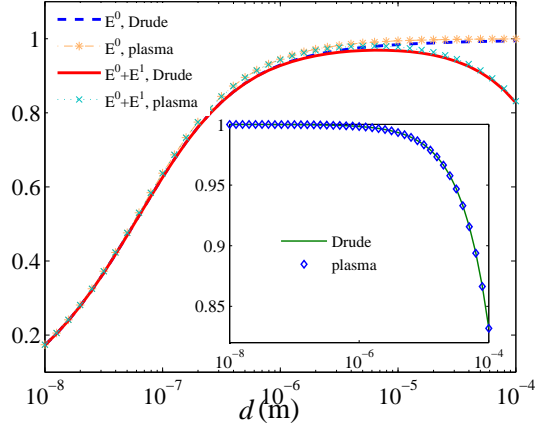} \caption{\label{f6} The   leading order term and the sum of the leading order and next-to-leading order terms of the Casimir interaction energy normalized by $ E_{\text{Cas}}^{\text{PFA,PC}}$.   The inset shows the ratio of the latter to the former.}\end{figure}

 \begin{figure} 
\epsfxsize=0.45\linewidth \epsffile{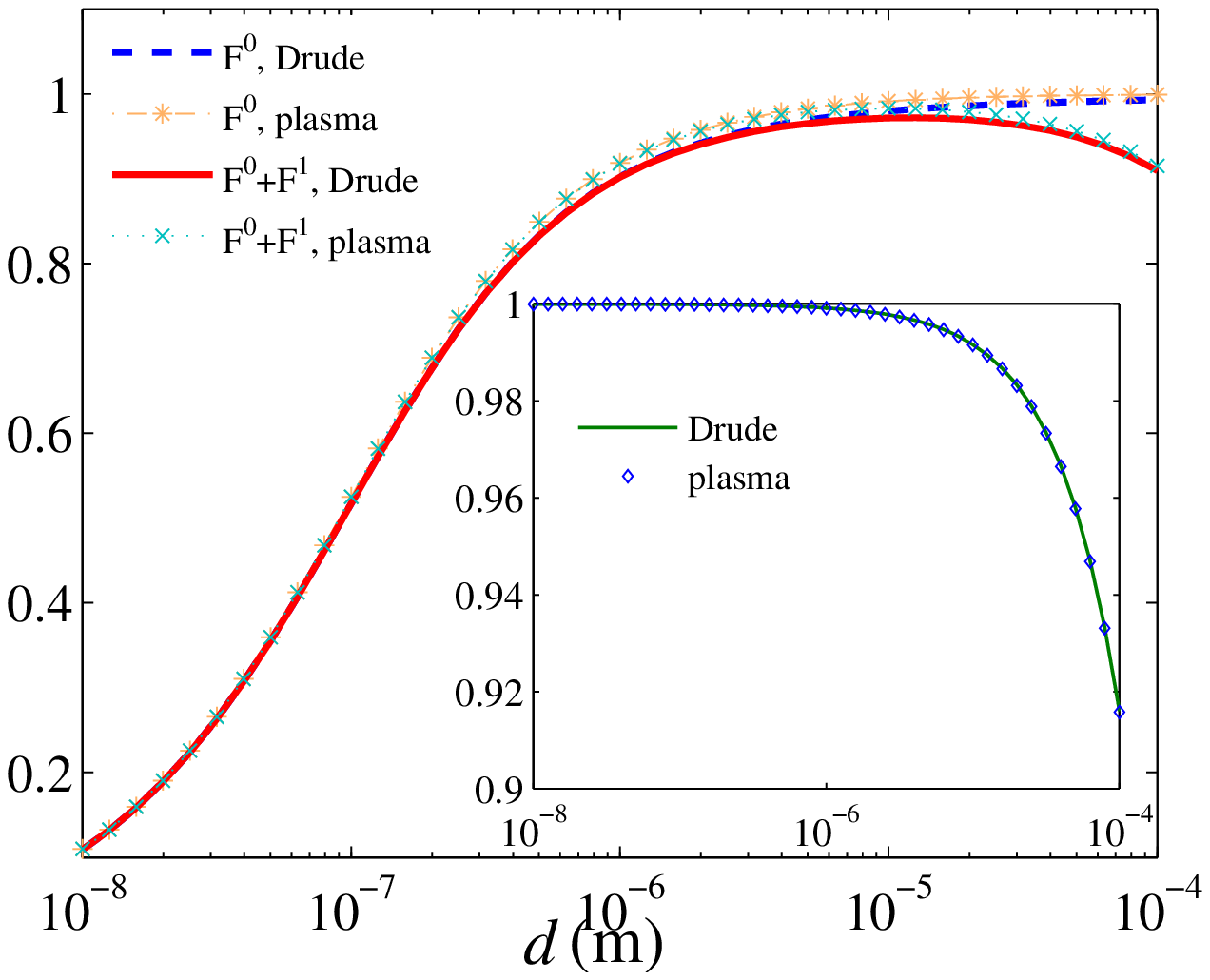} \caption{\label{f7} The   leading order term and the sum of the leading order and next-to-leading order terms of the Casimir force normalized by $ F_{\text{Cas}}^{\text{PFA,PC}}$.   The inset shows the ratio of the latter to the former.}\end{figure}

 \begin{figure}[h]
\epsfxsize=0.45\linewidth \epsffile{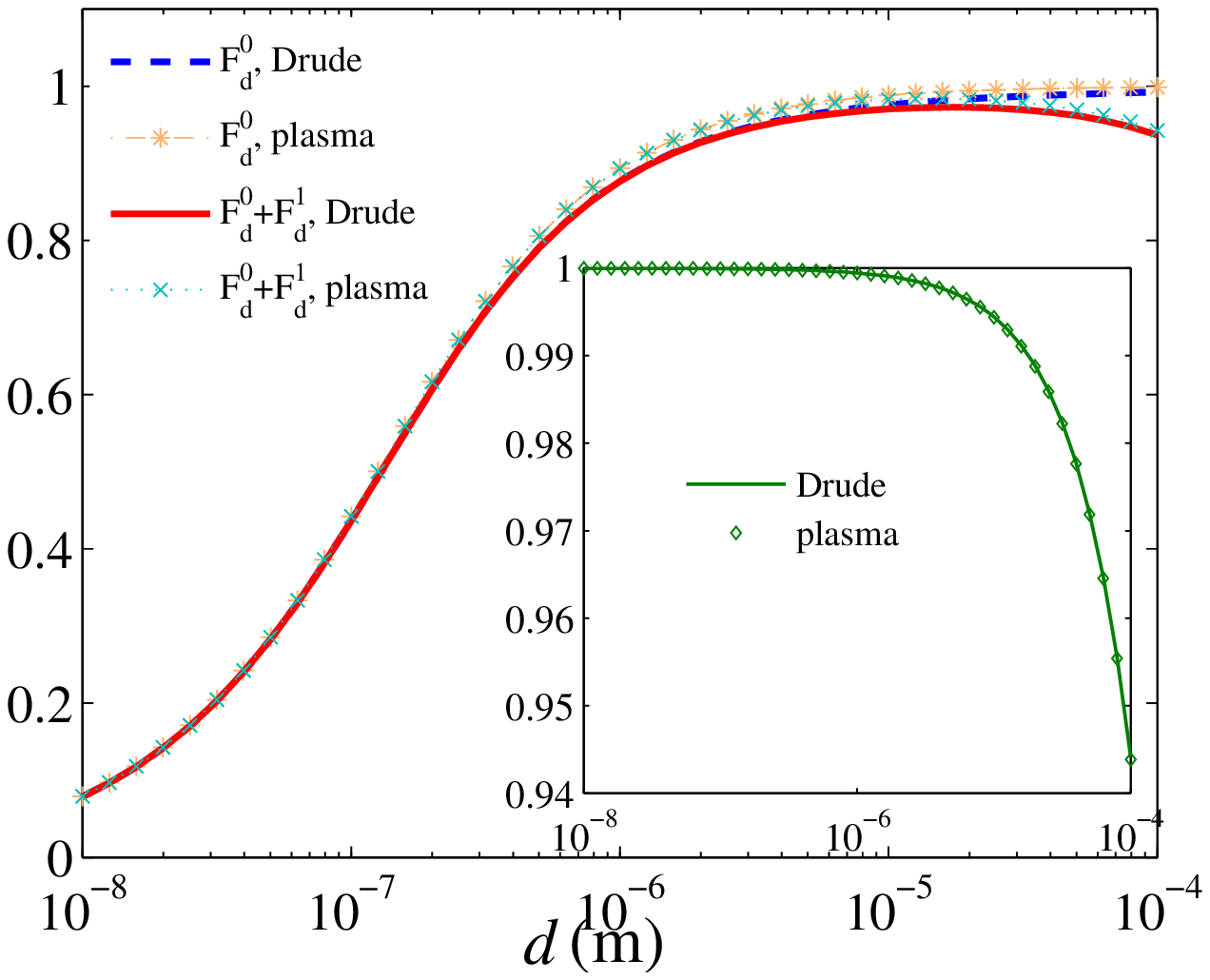} \caption{\label{f8} The   leading order term and the sum of the leading order and next-to-leading order terms of the force gradient normalized by $ {\pa F_{\text{Cas}}^{\text{PFA,PC}}}{\pa d}$.   The inset shows the ratio of the latter to the former.}\end{figure}

\begin{figure}
\epsfxsize=0.45\linewidth \epsffile{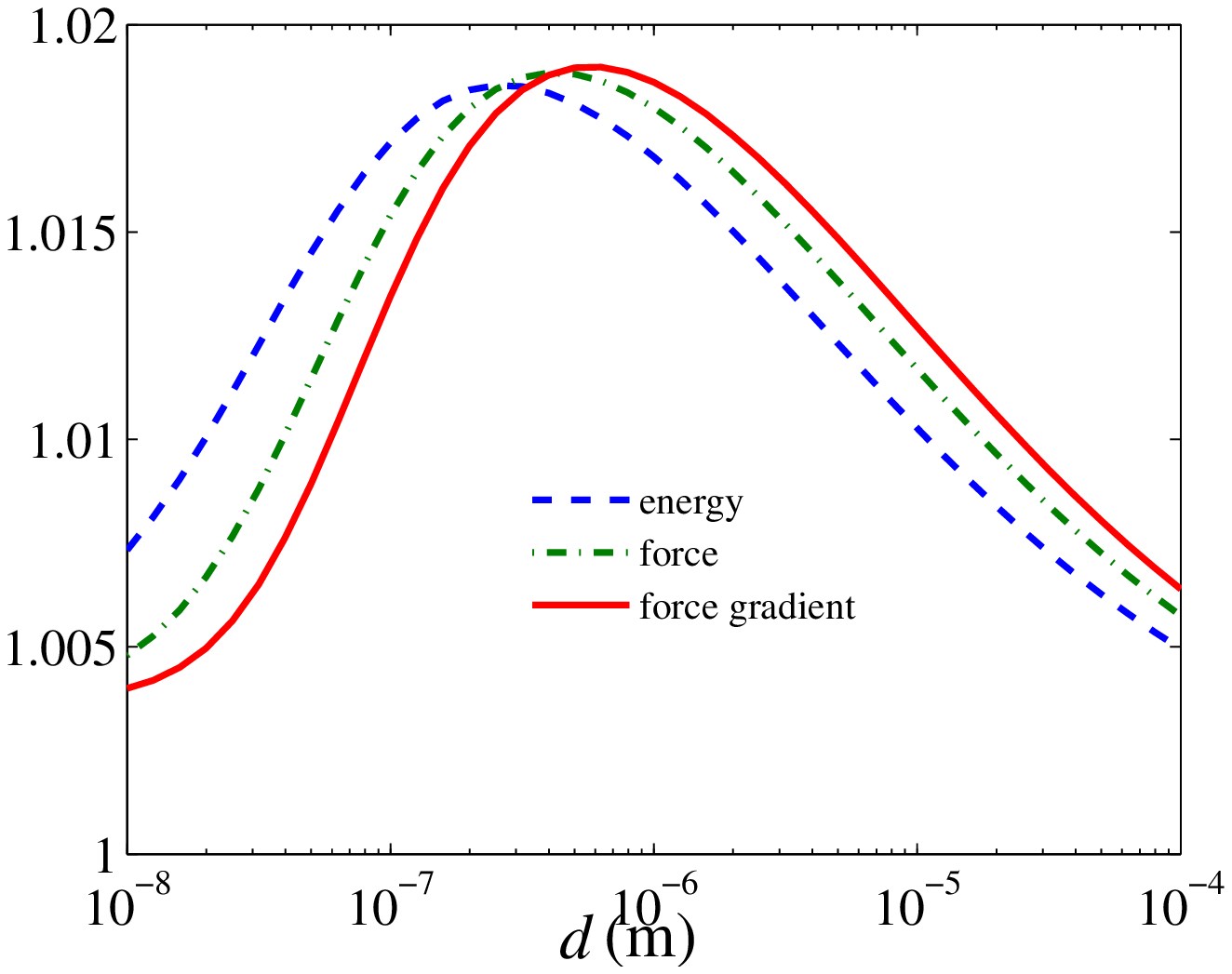}  \caption{\label{f11} The ratio of the plasma model to the Drude model for the sum of the leading order and next-to-leading order terms.    }\end{figure}

In Figs. \ref{f6}, \ref{f7} and \ref{f8}, we plot the leading order term, the sum of the leading order and next-to-leading order terms of the Casimir interaction energy, Casimir force and force gradient, normalized respectively by $ E_{\text{Cas}}^{\text{PFA,PC}}, F_{\text{Cas}}^{\text{PFA,PC}}$ and $\pa F_{\text{Cas}}^{\text{PFA,PC}}/\pa d$, as a function of $d$ for $d$ between 10nm and 100$\mu$m. Both the Drude model and the plasma model are plotted on the same graph to show the comparison.  To get a better picture, we plot the ratio of the plasma model to the Drude model for the sum of the leading order and next-to-leading order terms in Fig. \ref{f11}. From the figure, we notice that if the plasma model is used instead of the Drude model, the error is at most 2\%.

\begin{figure}[h]
\epsfxsize=0.32\linewidth \epsffile{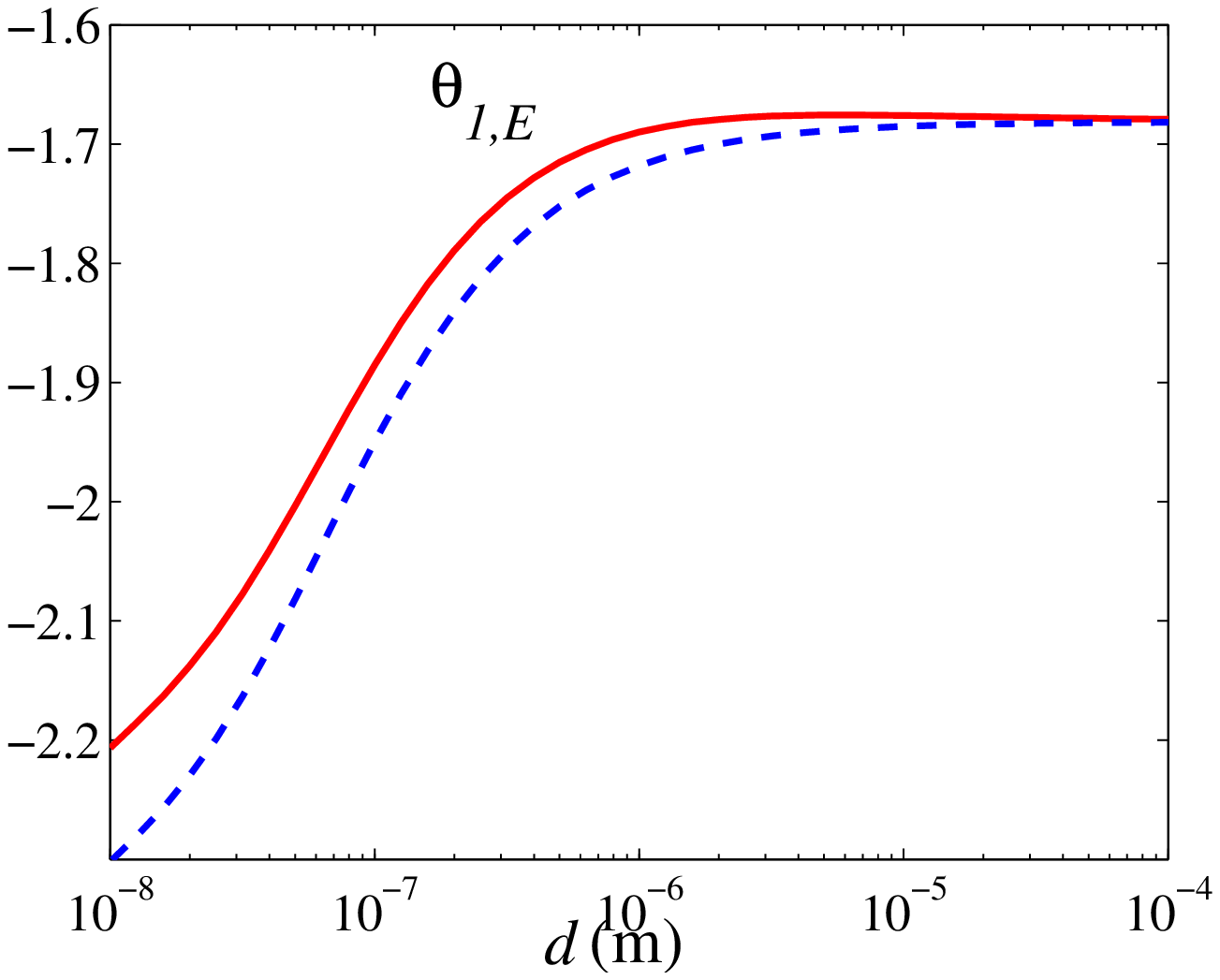} \epsfxsize=0.32\linewidth \epsffile{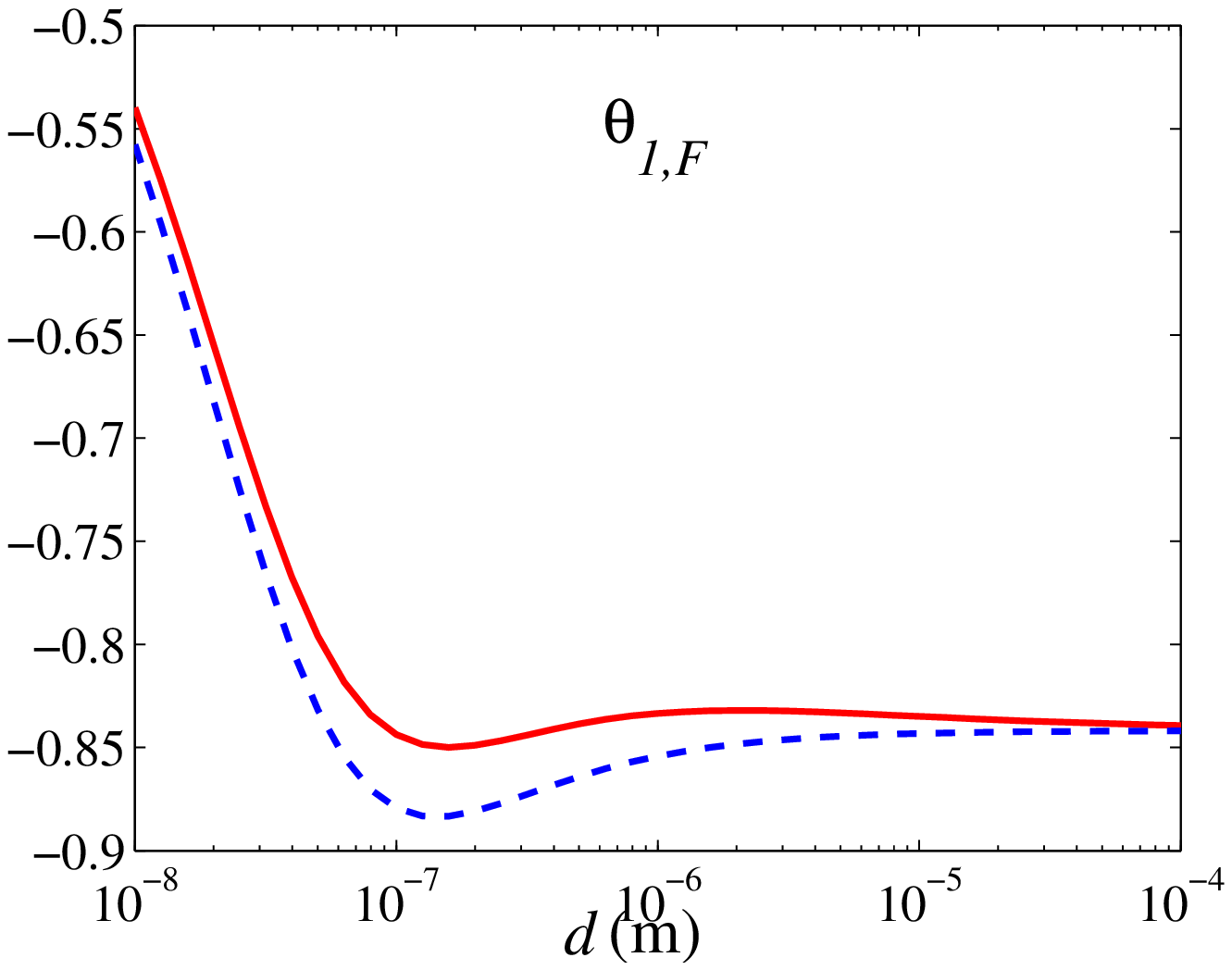} \epsfxsize=0.32\linewidth \epsffile{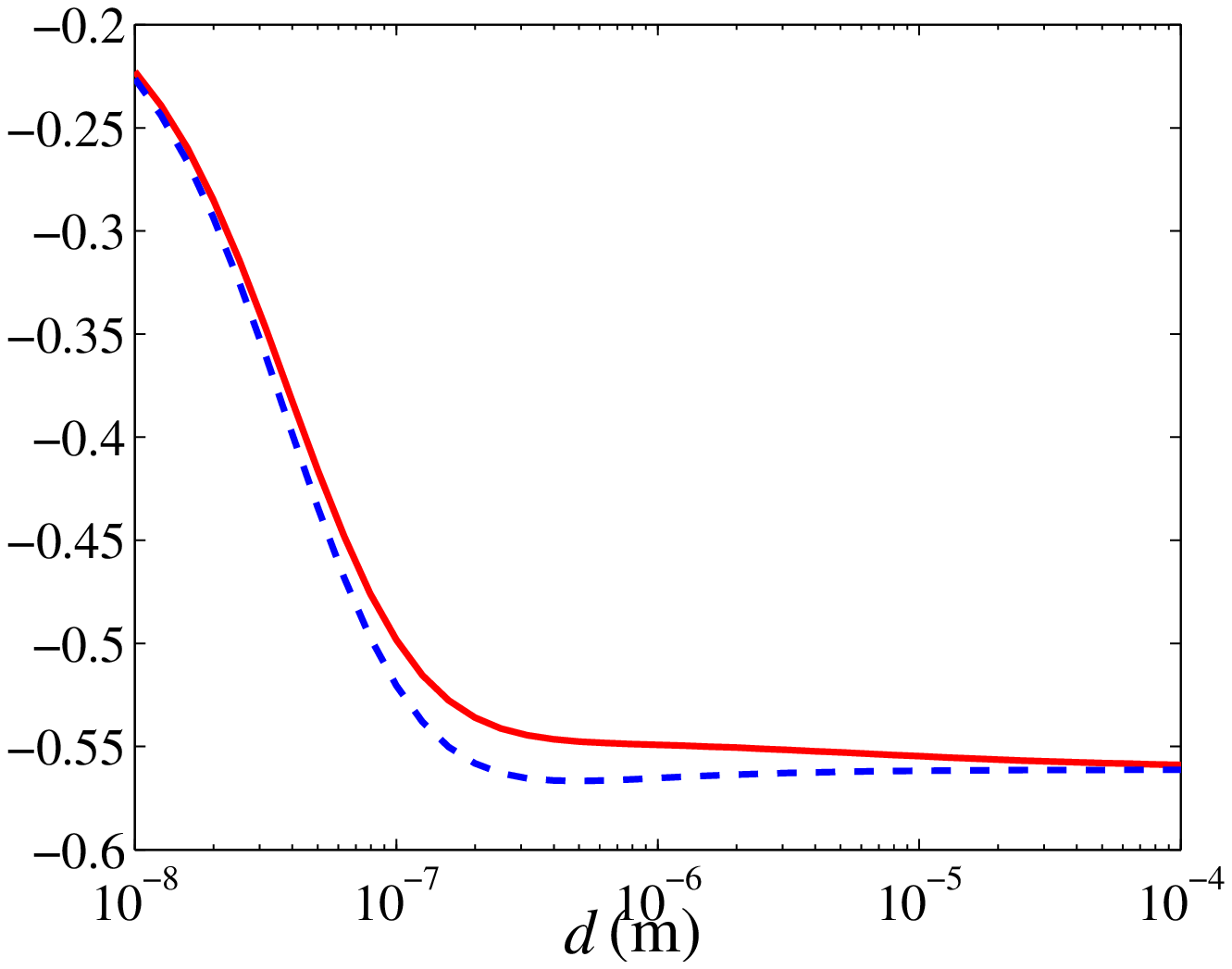} \caption{\label{f9}  $\theta_{1,E}(d)$, $\theta_{1,F}(d)$ and $\theta_1(d)$ computed using Drude  model (solid line), compared to those computed using plasma model (dashed line). }\end{figure}

In Figs. \ref{f9}, we plot $\theta_{1,E}, \theta_{1,F}$ and $\theta_1$ for Drude model and compare to that for plasma model. As for the plasma model, we notice that for the Drude model, as $d$ increases,   $\theta_{1,E}$, $\theta_{1,F}$ and $\theta_{1}$ tend respectively to the limiting values $\displaystyle 1/3-20/\pi^2 =-1.6931$, $\displaystyle  1/6-10/\pi^2=-0.8465 $ and $\displaystyle  1/9-20/(3\pi^2)=-0.5644 $, the corresponding values for perfect conductors. When $d$ is small, the deviations from these limiting values   are very significant. On the other hand,     $\theta_{1,F}$ and $\theta_1$ are also bounded from below.

\begin{figure}
\epsfxsize=0.45\linewidth \epsffile{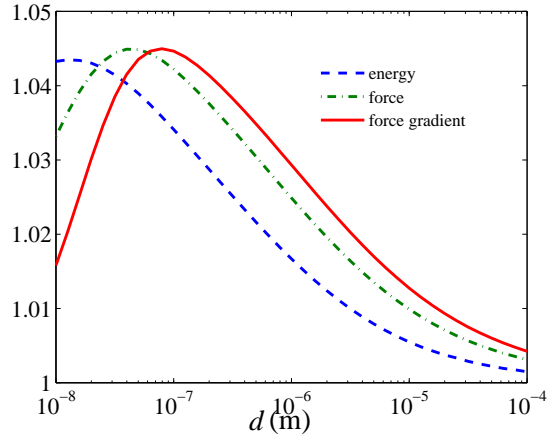}  \caption{\label{f12} The ratio of the plasma model to the Drude model for $\theta_{1,E}, \theta_{1,F}$ and $\theta_1$.  }\end{figure}

The ratios of the plasma model to the Drude model for $\theta_{1,E}, \theta_{1,F}$ and $\theta_1$ are plotted in Fig. \ref{f12}. From the figure, we find that if the  plasma model is used instead of the plasma model, the error is not more than 4.5\%.

\section{Conclusion}
Starting from the functional determinant representation of the Casimir interaction energy, we have used the perturbation method to obtain analytically the leading order and next-to-leading order terms of the Casimir interaction energy, Casimir force and force gradient for the interaction between a sphere and a plate. The results are written as double integrals over functions of the dielectric permittivities of the objects, and are hence general.
The leading order terms are shown to equal to that predicted by proximity force approximation. The results on the next-to-leading order terms are new, and they complement those obtained in \cite{12} using derivative expansion.

With given dielectric permittivities of the sphere and the plate, the double integrals representing the leading order and next-to-leading order terms can be computed numerically, and this is demonstrated for a gold sphere in front of a gold plate, where both plasma and Drude models are used for the dielectric functions of gold. It is observed that even at $d/R\sim 0.1$, the next-to-leading order term would contribute a correction to the leading order term of about 10\%. Of particular interest is the ratio of the next-to-leading order term divided by $d/R$ to the leading order term, denoted by $\theta_1$. It is found that when $\omega_p d/c$ is large enough, $\theta_1(d)$ tends to the corresponding limiting value for perfect conductors. However, when $\omega_p d/c$ is small, the deviation from the limiting perfect conductor value is significant. This signifies that in the nano range, we cannot model real metals by perfect conductors.

A comparison between plasma model and Drude model shows that their difference is below 2\% for the sum of the first two leading order terms and below 4.5\% for the values of $\theta_1(d)$. In fact, this small difference is expected at zero temperature.
In this work, we haven't considered the thermal effect. When $d$ is small enough such that $2\pi k_BTd/\hbar c$ is $\ll 1$, thermal effect can be neglected. For example, when $T=300$K, thermal effect can be neglected when $d\ll $ 1$\mu$m. Nevertheless, it would be interesting to examine the behavior of the Casimir interaction at room temperature and its interplay with material properties. This would be addressed in a forthcoming work.

\end{document}